\documentclass[10pt,journal,compsoc]{IEEEtran}
	\usepackage{afterpage}

	\usepackage{cite}
	\ifCLASSINFOpdf
	\else
	\fi
	\newcommand\T{\rule{0pt}{2.6ex}}       
\newcommand\B{\rule[-1.2ex]{0pt}{0pt}} 
	\usepackage{enumitem}
	\usepackage{soul}
	
	\usepackage{cite}
	\usepackage{amssymb,amsmath,caption}
\usepackage[hang,flushmargin]{footmisc}
\usepackage{lipsum}
\makeatletter
\newcommand{\algorithmfootnote}[2][\footnotesize]{%
  \let\old@algocf@finish\@algocf@finish
  \def\@algocf@finish{\old@algocf@finish
    \leavevmode\rlap{\begin{minipage}{\linewidth}
    #1#2
    \end{minipage}}%
  }%
}
	\usepackage{soul,color}
	\usepackage{hyperref}

\usepackage{subfig}
	
	\usepackage{ragged2e}
	
	\usepackage{amsmath,varwidth,array,ragged2e}
	\usepackage{multirow}

	\usepackage{graphicx}
	
	\usepackage{tabularx}
\usepackage{booktabs}
\usepackage{longtable}
\usepackage{multirow}
\usepackage{colortbl}
\usepackage{caption}
\usepackage{hhline}
\usepackage{tabularx,colortbl}
\usepackage{mathtools}
\usepackage{eqnarray,amsmath}
\usepackage{amsmath}
\usepackage{hyperref} 
\usepackage{cite}
\usepackage{amsmath,amssymb,amsfonts}
\usepackage{algorithmic}
\usepackage{graphicx}
\usepackage{enumitem}
\usepackage[utf8]{inputenc}
 \usepackage{subfig}
\usepackage{graphicx}
\usepackage{comment}
\usepackage{makecell}
\usepackage{wasysym}
\usepackage{mathrsfs}
\usepackage{amsthm}
\usepackage{bm}
\usepackage{soul}
\usepackage[export]{adjustbox}

\DeclareMathOperator*{\argmax}{argmax}
\DeclareMathOperator*{\argmin}{argmin}
\DeclarePairedDelimiter{\abs}{\lvert}{\rvert}

\usepackage[ruled,linesnumbered]{algorithm2e}
\definecolor{seagreen}{rgb}{0.18, 0.55, 0.24}
\SetAlFnt{\small\color{black}\tt}
\LinesNumbered

\usepackage{amssymb}
\usepackage{pifont}

\makeatletter
\newcommand\notsotiny{\@setfontsize\notsotiny\@vipt\@viipt}
\makeatother

\makeatletter
\newcommand*\bigcdot{\mathpalette\bigcdot@{.5}}
\newcommand*\bigcdot@[2]{\mathbin{\vcenter{\hbox{\scalebox{#2}{$\m@th#1\bullet$}}}}}
\makeatother

\usepackage{amssymb}
\usepackage{pifont}

	\usepackage{mathtools}  
	\usepackage{mathtools}
	\usepackage{amsmath}
\usepackage{amsthm}
\usepackage{amssymb}

	\usepackage{array}
	
	\usepackage{amsmath}
\usepackage{amsthm}
\usepackage{amssymb}
\usepackage[export]{adjustbox}
\usepackage{verbatim}
\usepackage{graphicx}

	\usepackage{threeparttable}

	\usepackage[table]{xcolor}
	\usepackage{tikz}
	\usetikzlibrary{shapes,arrows}
	
	\tikzstyle{line}=[draw] 
	
	\usepackage{amsfonts}
\usepackage{mathrsfs}

		\usepackage[export]{adjustbox}
	\usepackage{verbatim}
	\usepackage{graphicx}
	\usepackage{caption}

\begin{document}
	%
	\title{B-Ride: Ride Sharing with Privacy-preservation, Trust and Fair Payment atop Public Blockchain}
	\author{Mohamed Baza, Noureddine Lasla,~Mohamed~Mahmoud~(\IEEEmembership{Member~IEEE}),\\Gautam Srivastava~(\IEEEmembership{Senior~Member,~IEEE}), and~Mohamed~Abdallah~(\IEEEmembership{Senior~Member,~IEEE})
	\IEEEcompsocitemizethanks{\IEEEcompsocthanksitem Mohamed Baza, and Mohamed Mahmoud are with the Department of Electrical \& Computer
	Engineering, Tennessee Tech University, Cookeville, TN 38505 USA.\protect\\
		E-mails:\enskip{}\href{mailto:mibaza42@students.tntech.edu}{mibaza42@students.tntech.edu},
		 \href{mmahmoud@tntech.edu}{mmahmoud@tntech.edu}.

	\IEEEcompsocthanksitem Noureddine Lasla and Mohamed Abdallah are with the division of Information and Computing Technology, College of Science and Engineering, HBKU, Doha, Qatar.
	E-mails:\enskip{}
		\href{mailto:lasla.noureddine@gmail.com}{nlasla@hbku.edu.qa},
	\href{mailto:moabdallah@hbku.edu.qa}
	{moabdallah@hbku.edu.qa}

	\IEEEcompsocthanksitem Gautam Srivastava is with Department of Mathematics and Computer Science, Brandon University, Manitoba, Canada.
	E-mail:\enskip{}\href{mailto:SRIVASTAVAG@BrandonU.CA}
	{SRIVASTAVAG@BrandonU.CA}

	}
	

	}

\IEEEtitleabstractindextext{
	\begin{abstract}
	
 Ride-sharing is a service that enables drivers to share trips with other riders, contributing to appealing benefits of shared travel cost and reducing traffic congestion. However, the majority of existing ride-sharing services rely on a central third party to organize the service, which make them subject to a single point of failure and  privacy disclosure concerns by  both internal and external attackers. Moreover, they are vulnerable to distributed denial of service (DDoS) and Sybil attacks launched by malicious users and external attackers. Besides, high service fees are paid to the ride-sharing service provider. In this paper, we propose a decentralized ride-sharing service based on public Blockchain, named B-Ride. B-Ride enables drivers to offer ride-sharing services without relying on a trusted third party. Both riders and drivers can learn whether they can share rides while preserving their trip data, including pick-up/drop-off location, departure/arrival date and travel price. However, malicious users exploit the anonymity provided by the public blockchain to submit multiple ride requests or offers, while not committing to any of them, in order to find a better offer or to make the system unreliable. B-Ride solves this problem by introducing a time-locked deposit protocol for a ride-sharing by leveraging smart contract and zero-knowledge set membership proof. In a nutshell, both a driver and a rider have to show their good will and commitment by sending a deposit to the blockchain. Later, a driver has to prove to the blockchain on the agreed pick-up time that he/she arrived at the pick-up location on time. To preserve rider/driver privacy by hiding the exact pick-up location, the proof is performed using zero-knowledge set membership proof. Moreover, to ensure fair payment, a pay-as-you-drive methodology is introduced based on the elapsed distance of the driver and rider. In addition, we introduce a reputation model to rate drivers based on their past behaviour without involving any third-parties to allow riders to select them based on their history on the system.  Finally, we implement our protocol and deploy it in a test net of Ethereum. The experimental results show the applicability of our protocol atop existing real-world blockchains.

	\end{abstract}
	
	
	}

	\maketitle

	\IEEEdisplaynontitleabstractindextext


	%

	\IEEEpeerreviewmaketitle

	\IEEEraisesectionheading{\section{Introduction}\label{sec:introduction}}

	\IEEEPARstart{O}{ver} the last few years, ride-sharing services (RSSs) have been emerging  as an alternative transportation services that allow people to use personal cars more wisely. In RSSs, a driver shares his vacant car seats with other riders. Ride sharing has several benefits to the individual and the community at large including increasing occupancy rates, sharing travel costs, extending social circles, and reducing both fuel consumption and air pollution~\cite{schrank20152014, sanchez2016co}. Across the world, many providers that offer online ride-sharing services such as Flinc, UberPool, Lyft Line and Blablacar have emerged.
	According to~\cite{R12}, the ride sharing market is projected to reach USD 218 billion by 2025.


A ride-sharing service can find the drivers and riders who can share rides  in matching drivers using their ride offers (i.e., planned trips)
and ride requests (i.e., desired trips). To enable ride-sharing service, users
(i.e., drivers and riders) have to share with a service provider
the trip information, including departure time, location, and destination. The service provider works as a middleman to facilitate the communication between the system users and usually charges a commission for each successful shared ride.
However, running the service by a central server, makes the system vulnerable to a single point of failure and attacks~\cite{baza2019blockchain}. If the security of the service provider is compromised, the service can be interrupted and the data can be disclosed, altered, or even deleted. For instance, Uber has witnessed a tremendous data leakage of $57$ million customers and drivers for more than a year. Uber has paid $148$ million just to settle an investigation to its data breach. 
Similarly, in April 2015, due to hardware failure in Uber China, a service outage occurred and passengers were not able to complete their orders at the end of services~\cite{china}. 
In addition, in order to maximize their own benefits, most ride-sharing service providers impose a high service fee that can reach up to 20\%  \cite{li2018crowdbc,baza2018blockchain}. 

In contrast to the traditional client-server model, Blockchain\footnote{Throughout this paper, a blockchain refers to permissionless/public blockchain that lets any interested party to participate and leave, as opposed to the less ambitious way of having blockchain atop permissioned parties.} is a verifiable, immutable and distributed ledger that allows mistrusting entities to
transact with each other without relying on a central third party. Blockchain is a transparent  data structure that is organized as a chain of blocks and managed by a network of computers, called miners,
running a peer-to-peer (P2P) protocol. Each block contains a set of transactions that are committed by network peers
according to a predefined consensus algorithm~\cite{kosba2016hawk,parksmarnet}. Blockchain was first introduced as a distributed cryptocurrency that enables the transfer of electronic cash without the intervention of banks. Since then it has evolved beyond that to support the deployment of more general-purpose distributed applications. This concept has been introduced by Vitalik Buterin and is called smart-contracts or decentralized autonomous organizations \cite{wood2014ethereum}. A smart-contract can be described as an autonomous computer program running on blockchain network. This program acts as a contract whose terms can be pre-programmed with the ability of self-executing and self-enforcing itself without the need for trusted authorities~\cite{christidis2016blockchains,parkccnc}.

In this paper, we propose a blockchain-based ride-sharing system using smart-contracts to mitigate the single point of failure issues presented in classical client-server architectures.
However, besides being completely distributed and transparent, the openness of blockchain leads to a potential privacy concern where the data can be publicly accessible. Despite the use of anonymous authentication, this is not sufficient to protect the privacy of the end users. For instance, by tracking the activity of a driver or rider, an attacker with little background knowledge of that user can figure out all his location traces \cite{kosba2016hawk}. Moreover, because in public blockchains, anyone can join and transact in the network anonymously, malicious users can disturb the blockchain-based ride-sharing service by sending, for instance, multiple requests/offers while not committing to any of them. Therefore, it is required to keep track of drivers' behaviours and build a reputation system that helps a rider to select with confident an appropriate driver for his ride request. Subsequently, in order to decentralize ride-sharing services in a meaningful way, the privacy concern with respect to ride-sharing needs to be carefully evaluated and addressed. This mainly requires resolving two conflicting objectives, i.e., $(i)$ the desire to have a \textit{transparent} system while protecting users \textit{privacy}, and $(ii)$ ensure \textit{accountability} while being \textit{anonymous}.

Motivated by the above challenges, we propose B-Ride a \textbf{B}lockchain-based \textbf{Ride} sharing service that is privacy preserving while also establishing trust between drivers\footnote{Hereafter we use the term "driver" to refer to individuals or companies that own vehicles or buses that can be used in a ride-sharing services.} and riders. \textit{To the best of our knowledge, this work is the first to employ ride-sharing services atop open and permissionless blockchains}. B-Ride aims to remove intermediaries between riders and drivers and make use of blockchain and smart contracts vetting to the future of ride-sharing services.
 
 Our main contributions and the challenges the paper aims to address can be summarized as follows:

\begin{enumerate}

\item  A blockchain based system is proposed to realize decentralized ride-sharing services. To preserve riders' trip privacy, we use \textit{cloaking}, so a rider posts a cloaked pick-up and drop-off location as well as pick-up time. Then, interested drivers use off-line matching technique to check if the request falls on his cloaked route and then send the exact trip data encrypted with riders' public key. Then, a rider can select the best-matched driver to share a trip based on some heuristics. This acts as a distributed auction that is handled through the blockchain to ensure transparency.

\item To ensure trust between a rider and a selected driver, we propose a time-locked deposit protocol for ride-sharing services based on the zero-knowledge set membership ~\cite{camenisch2008efficient}. The core idea is to define \textit{claim-or-fine} methodology that works as follows; $(i)$ A rider has to post a smart contract with a deposit budget as proof of accepting a driver's offer as well as a \textit{set} of different obfuscated locations. $(ii)$ The selected driver should also deposit a budget to the contract as a commitment to his offer. $(iii)$ Upon arrival at the pick-up location, the driver acts as (\textit{a prover}) and sends a \textit{proof to pick-up location} to the blockchain. Specifically, the driver proves that the pick up location falls in a predefined set of cells. $(iv)$ Finally, a smart contract acts as a (\textit{verifier}) by checking the proof in a zero-knowledge manner and then assigns rewards to driver in case of valid proof or fine the driver in case of invalid or if no proof is sent before the agreed pick-up time.

\item Also, we propose a method to ensure fair payment in a trust-less manner between the driver and rider. A driver needs to send at a regular interval an elapsed distance to the rider who authenticates it by signing it using his private key. Then, once the rider provides a \textit{proof-of-elapsed-distance} (i.e., the elapsed distance and driver' signature on it), the smart-contract transfers the fare to the driver. In this way, the driver gets paid as he/she drive. Meanwhile, if the rider stops sending proofs to the blockchain, he/she can stop the trip immediately. Moreover, only elapsed distances are stored on the blockchain and no other sensitive information are leaked to the public.


\item Finally, B-Ride computes the \textit{reputation} of drivers based on their prior behaviours. Unlike, current centralized reputation approaches~\cite{wu2017sharing}, we develop a decentralized reputation management mechanism over blockchain that is executed in a self-enforcing manner once a predefined set of conditions are met. Specifically, in B-Ride, each driver has two reputation indices; $(i)$ The first score increases every time a driver sends a valid proof of his arrival to the pick-up location. $(ii)$ The second score increases upon the completion of each trip. Based on the two indices, each driver will have a trust value in B-Ride that will be used by riders to select good drivers for their trips. Our reputation mechanism makes economic incentive for drivers to behave correctly, otherwise they will will not be selected by anyone.

\item To evaluate B-Ride, we implement it on top of Ethereum, a real-world public blockchain platform. Intensive experiments and performance evaluations are conducted in an Ethereum test network. 

\end{enumerate}

 The rest of the paper is organized as follows. In Section~\ref{Preliminaries}, we discuss preliminaries used by this research work. We describe the network and threat models, followed by the design goals of our system in Section~\ref{sec: models}. A detailed description of our system is presented in Section~\ref{sec:ProposedScheme}. Performance evaluations are presented in Section.~\ref{comm overhead}. We present the security, privacy, and computation complexity analysis of our scheme in Section~\ref{sec:analysis}. Section~\ref{Related} discusses the previous research work. Finally, we give concluding remarks in Section~\ref{conclusion}, followd by an acknowledgement in Section~\ref{Ack}.

\section{PRELIMINARIES}
\label{Preliminaries}
In this section, we present the necessary background on blockchain, smart contracts and some cryptosystems that are used in this paper. The main notations used in this paper are given in Table 1.

\subsection{Blockchain and Smart Contracts}
\label{sec: blockchain and contract}


Blockchain serves as the enabling technology for emerging
cryptocurrencies like Bitcoin~\cite{nakamoto2008Bitcoin} to make a peer-to-peer exchange of value without relying on a third party.
A blockchain is an immutable, distributed, and append-only
data structure created by a sequence of blocks which are
chronologically and cryptographically tied together~\cite{wood2014ethereum}. Typically, blockchain is a network composed
of a set of nodes, named \textit{miners or validators}, which are responsible for keeping trustworthy records of all transactions using a consensus algorithm in a trust-less environment.
More importantly, blockchain enables the essence of smart
contracts which can be defined as programs that every
blockchain node  runs  and updates their local replicas
according to the execution results without any
intervention from a third party.

The unique features of blockchains are: $(i)$ \textit{Transparency} since
transactions stored on the blockchain are visible to all
participants in the network. $(ii)$ \textit{Liveness} since all participants
can reach the same blockchain while new blocks with valid
transactions will continue to be added~\cite{kosba2016hawk}. $(iii)$ \textit{Eventual consensus} because transactions stored on the blockchain should be validated and a secure consensus protocol run by all
participants to agree on its global state~\cite{bonneau2015sok}. $(iv)$ \textit{Blockchain address or Pseudonym}, that is referred to the message sender in the blockchain. In practice, a blockchain address is bound to the hash of a public key~\cite{guo2018blockchain,guo2019access}. Specifically, the security of digital signatures assures that no one can send messages in the name of a blockchain address, unless it
knows the corresponding secret key. Likewise, the smart contract deployed in the blockchain also has a unique address, such that anyone can call the contract to be executed.

\subsection{Notations}

 Let $PG$ be a pairing group generator that on input $1^{k}$ outputs descriptions of multiplicative groups $\mathbb{G}_{\mathrm{1}}$ and $\mathbb{G}_{\mathrm{T}}$ of
prime order $p$ where $|p|=k$. The generated groups are such that there exists an admissible bilinear map $e : \mathbb{G}_{1} \times \mathbb{G}_{1} \rightarrow \mathbb{G}_{\mathrm{T}}$
meaning that (1) for all $a, b \in \mathbb{Z}_{p}$ it holds that $e\left(g^{a}, g^{b}\right)=e(g, g)^{a b}$; (2) $e(g, g) \neq 1$ and (3) the bilinear map is efficiently computable. $\mathcal{H}$ is a
collision-resist hash function that maps strings of arbitrary
length to $\mathbb{Z}_{p}$.We denote $u \in_{R} \mathbb{Z}_{p}$ as randomly choosing a number from $\mathbb{Z}_{p}$.

	\begingroup\renewcommand{\caption}[1]{}%
			\begin{figure}[!t]
		\centering
		\includegraphics[width=1\linewidth]{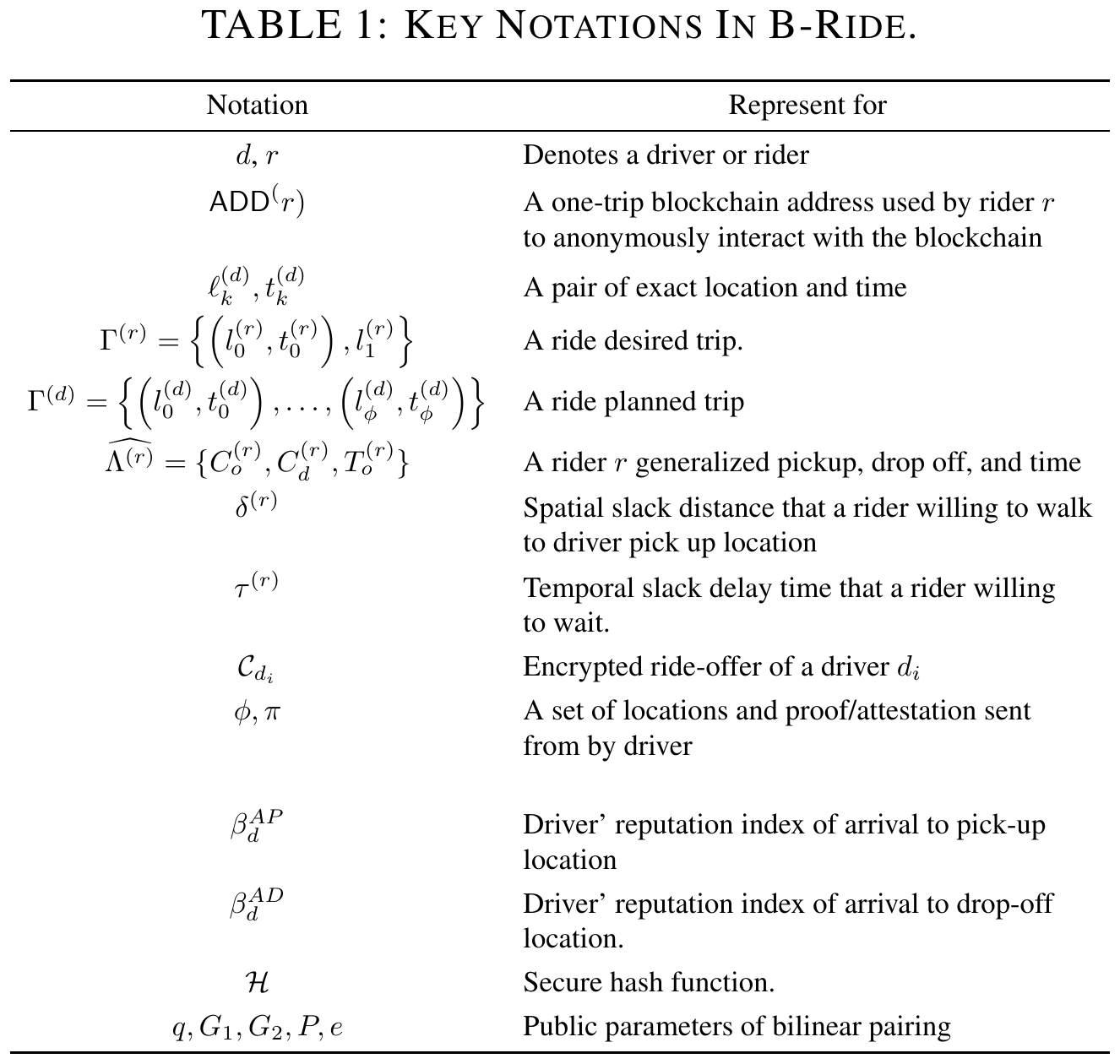}
		\vspace{0mm}
		
		\caption{Illustration of the bidding and selection phase in B-Ride. }
	\end{figure}
\endgroup

\subsection{Zero Knowledge Set Membership Proof (ZKSM)}
\label{ZKSM}
A set membership proof enables a prover to prove, in a
zero-knowledge way, that a secret value lies in a given public
set. The set can perhaps be a list of cities or clubs. Typically, such proofs can be used, for example, in the context of electronic voting, where a voter is required to prove that his secret vote belongs to the set that contains all possible candidates. We are going to use Camenisch and Stadler~\cite{camenisch1997efficient} notation for proofs of knowledge: 

\begin{equation}
\boldsymbol{PK}(\delta, \gamma): Y=g^{\delta} h^{\gamma} \land ( \gamma \in \phi)  
\end{equation}

Where  $Y=g^{\delta} h^{\gamma}$ is a Pedersen commitment of the integer $\delta \in \mathbb{Z}_{p}$ using randomness $\gamma$. The above proof convinces the verifier
that the secret in the commitment $Y$ lies in the set $\phi$ without having to explicitly list $\phi$ in the proof. The set $\phi$ can be a common input to both prover and verifier. The Set membership proof can be instantiated in the discrete logarithm setting and made non-interactive with Fiat-Shamir heuristic. We refer the readers to~\cite{camenisch1997efficient} for the detailed construction.

The security guarantees are: $(i)$ \textit{soundness:} no prover can convince an honest verifier if he/she did not compute
the results correctly; $(ii)$ \textit{completeness:} if the statement is true, the honest verifier (the one following the protocol properly) will be convinced of this fact by an honest prover; $(iii)$ \textit{zero-knowledge:} the proof distribution can be simulated without revealing any secret state, i.e., no verifier learns anything other than the fact that the statement is true. 


\section{Network/Threat Models AND Design Goals}
\label{sec: models}
In this section, we describe the considered network and threat models. Also, we define the adversarial assumptions. 
\subsection{Network model}
 As depicted in Fig.~\ref{fig:Bride model}, the considered network model has the following entities.
 
\begin{enumerate}[label={}]
  
 \item   \textit{Blockchain.} At the heart of our system is the blockchain network that handles all the ridesharing transactions. We opt for a permissionless blockchain where everyone can use the system to either act as a driver or rider. The ridesharing business logic is defined in a smart-contract and executed by the blockchain network, as described in Section~\ref{sec: blockchain and contract}. The  blockchain is also used for peer-to-peer payment to allow the  exchange of currency between the different system users. Thus, we select Ethereum, which is the most popular open blockchain platform for smart-contracts, to implement and evaluate our proposed protocol.


\vspace{6pt}
    \item  \textit{Drivers and riders.} The rider is an entity that requests for ridesharing service by making a request to the system. The driver is an entity that wants to share his trip by posting an offer to riders' requests. Note that drivers and riders are not required to store a complete copy of the blockchain. Instead, they can run on top of light-weight nodes, which eventually enables them to interact with the network to send   transactions or read from the blockchain~\cite{lu2018zebralancer}. 
		\begin{figure}[!t]
		\centering
		\includegraphics[width=1\linewidth]{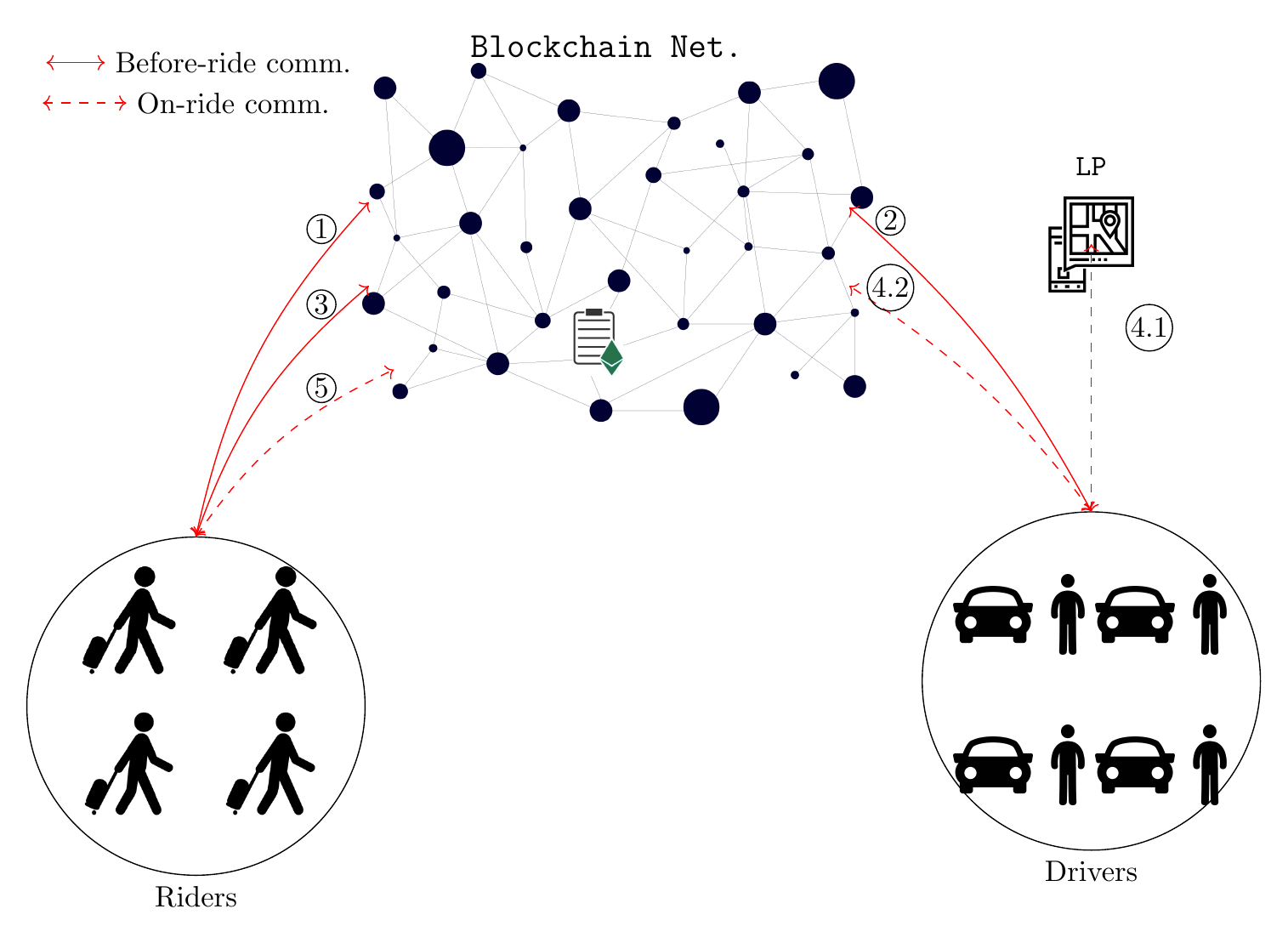}
		\vspace{-7mm}
		\caption{System architecture: (1) The rider publishes a ride request contract to the blockchain (2) Drivers sends their encrypted offers. (3) A rider selects the best matched offer and publish a time-locked contract. (4.1) and (4.2) Up on arrival, the driver sends a proof of arrival to pick-up and claim rider deposit (5) The rider publishes a payment contract that transfer the fare trip to the driver. }
		\label{fig:Bride model}
	\end{figure}
        
   \vspace{6pt}
    \item  \textit{Location Prover (LP):}  The role of LP is to ensure the authenticity of the  reported location (pickup) by the drivers. In our scheme, upon arrival of a driver to a given pickup location, he/she should send its current location to the blockchain through the LP. An LP could be, for instance,  roadside units that are already deployed to enable Vehicular Ad-hoc Networks (VANETs)~\cite{baza2019detecting}, where each RSU can confirm if the location provided by a driver falls in its coverage or not.

 \end{enumerate}

\begin{figure}[!t]
		\centering
		\includegraphics[width=.88\linewidth]{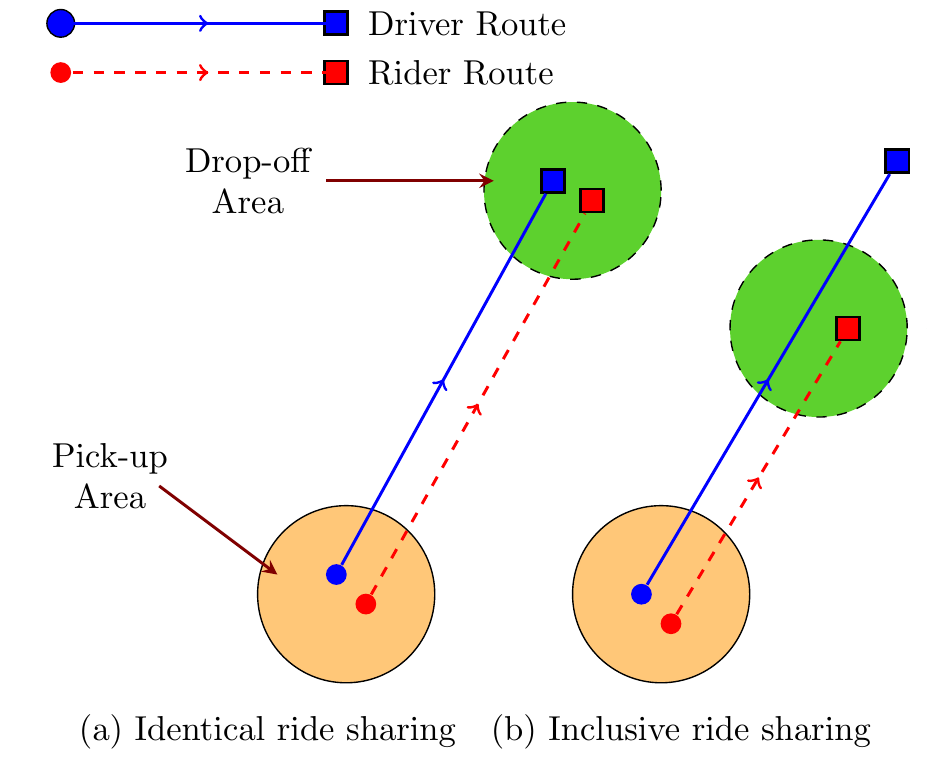}
		\vspace{0mm}
		
		\caption{Ride sharing cases under consideration in B-Ride.}
		\label{2}
\end{figure}

Ride sharing allows small tolerance in both locations and times of the requests/offers~\cite{furuhata2013ridesharing}. Thus, spatial and temporal slacks are introduced to capture the maximum additional distance $\delta$,  and time $\tau$ that  a rider or driver can tolerate for traveling, and  waiting, respectively. In this paper,  we consider the 
following ride sharing cases:
\begin{enumerate}
        \item \textit{Identical ridesharing.} The pick-up and drop-off locations of both a driver and rider are matched, as shown in Fig. \ref{2}.a.
        \item \textit{Inclusive ridesharing.} The destination of the rider lies on the driver's route. In this case, the driver must stop for drop-offs before reaching its final destination, as shown in Fig. \ref{2}.b.
\end{enumerate}


\subsection{Adversarial and threat Assumptions}
We assume that both internal and external adversaries could try to compromise the security of the ride sharing system. We follow the standard blockchain threat model presented in~\cite{kosba2016hawk}, where the blockchain is trusted for execution correctness and availability, but not for privacy. The smart-contract code is visible and checkable by anyone once it is deployed and is guaranteed to work as specified, free from tampering. Likewise, any data submitted and stored to the contract can be directly read by all parties of the system as well as any  external curious users. In addition, the following threats are also considered:
\begin{itemize}
    \item Global eavesdroppers can read all transactions recorded on the blockchain for the riders in order to learn their moving patterns, guess their locations at a specific time or even track them over the time.
	 \item A rider may perform a location cheating attack by reporting a false planned trip to the blockchain. The rider fraudulently does not commit to the request  whereas the victim driver has to travel a long way to pick up the rider. Likewise, a driver may unfairly match more riders while deliberately do not commit to this offers.
	 
	\item Cheating of fare payment. If a driver gets paid at the beginning of the trip, he/she may misbehave and does not complete the trip. Also, if a driver gets the trip fare at the end of trip, the rider may not be willing to pay the fare~\cite{over}.
\end{itemize}

\subsection{Design and Functionality Requirements}
Under the aforementioned system model and adversarial assumptions, our aim is to develop a ride-sharing system with the following design goals:
\begin{enumerate}
\item \textit{Resilience.} The proposed scheme should not rely on a central ride-sharing organizing entity, and no party in the system is completely trusted.

\item  \textit{Preserving riders' privacy.} The proposed scheme should preserve riders' privacy including their trip data, i.e., pick up/drop off. This can be satisfied if the following two conditions are achieved: $(i)$ None of the drivers/miners, except for the selected driver, learns the exact position of the rider or the associated driver. $(ii)$ A specific rider cannot be tracked over time.
  
 \item \textit{Ensuring trust between riders and drivers.} It is important to discourage any malicious behavior by both drivers and riders. This objective is challenging especially when relying on a public permissionless blockchain and also considering privacy concerns. 
 
 \item \textit{Ensuring fair payment.} The proposed ride sharing service should ensure fair payment in order to attract more users to the system. Hence, the payment should be done in a trust-less manner to protect honest drivers from dishonest riders and vice versa. 

\item \textit{Drivers reputation management.} The proposed scheme should keep track of drivers' behavior through a reputation system. Malicious drivers who may try to subvert the system, even irrationally, can be identified by low reputation scores. Therefore, drivers having low reputation scores will be distrusted and no one will be interested to interact with them.
\end{enumerate}
	
\section{Our Proposed Scheme: B-Ride}
\label{sec:ProposedScheme}
B-Ride consists of  the following six phases: trip data generation, bidding and selection, time-locked deposit protocol, fair payment and reputation management.

\begin{figure}[!t]
		\centering
		\includegraphics[width=.99\linewidth]{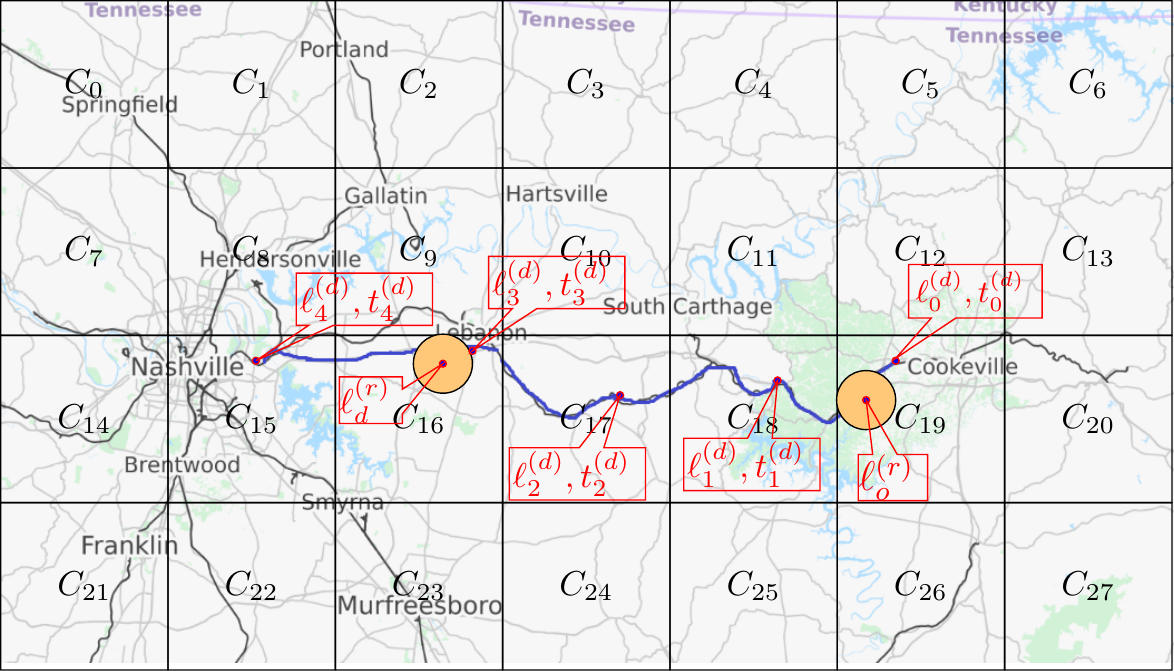}
		\vspace{0mm}

		\caption{Illustration of dividing the ride sharing area of interest into cells for the state of TN, USA. A driver $d$'s route, in blue, with 5 points, and pickup and drop-off of a rider $r$.}
		\label{fig1: map}
\end{figure}

\subsection{Trip Data Generation}
\label{trips}

\begin{figure*}[!t]
		\centering
		\includegraphics[width=.8\linewidth]{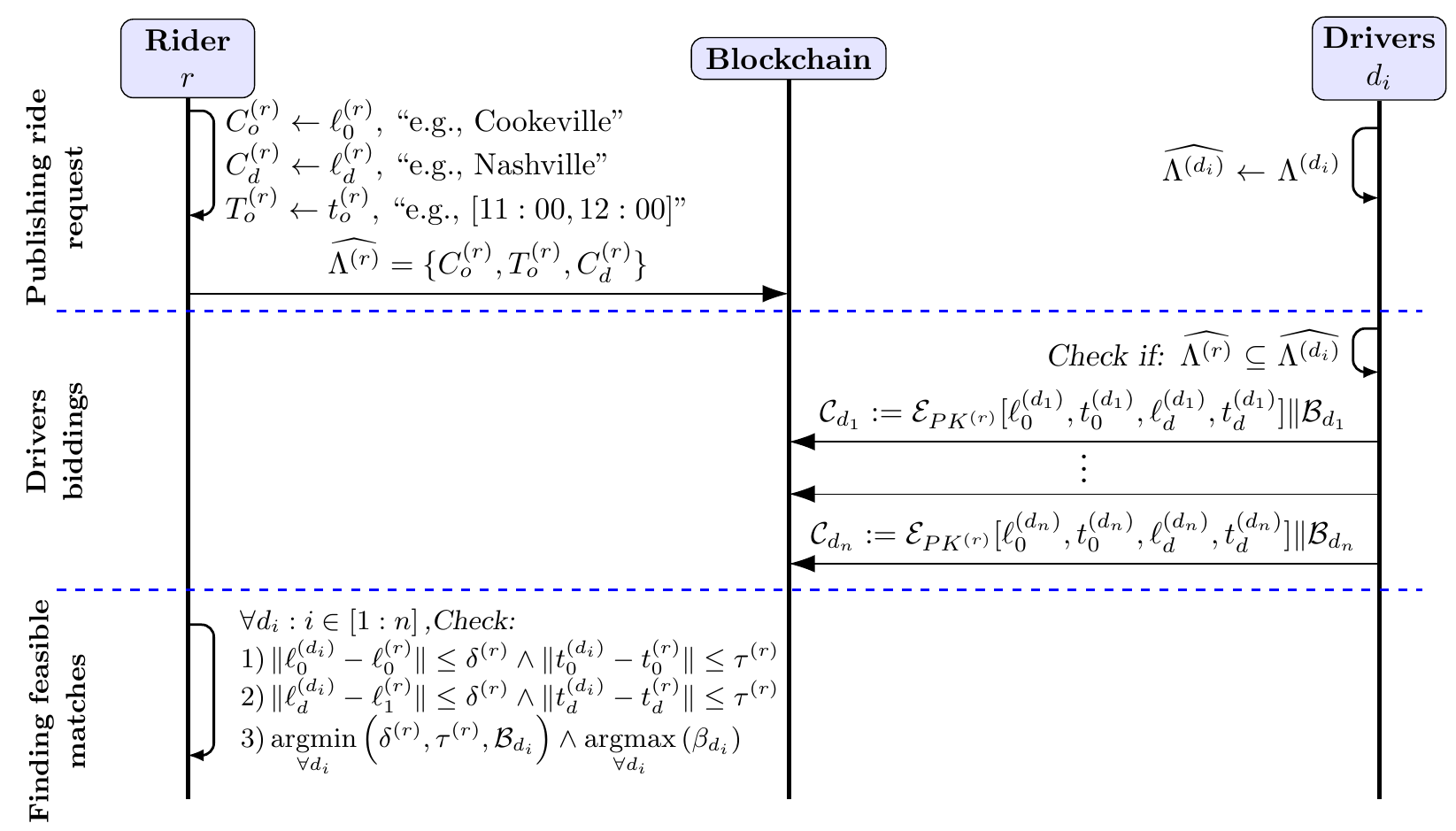}
		\vspace{0mm}
		
		\caption{Illustration of the bidding and selection phase in B-Ride.}
		\label{Bidding}
\end{figure*}

In this section, we discuss how the drivers/riders generate their trips data while preserving their privacy. Generalization technique~\cite{sanchez2016co} also known as spatial cloaking, is used for this purpose. The main idea of the spatial cloaking is to blur users' exact locations into cloaked regions for location obfuscation. Let's denote by $\mathcal{A}$ the ride sharing area (e.g., a state) that is sub-divided into a set of $n$ \textit{cells} $C=\{C_1, C_2, ... C_n\}$. Cells could be defined based on geographic area constraints such as  districts or neighborhoods in a city, uniform partitions in a map, etc. Therefore, instead of using the exact pick-up/drop-off location, riders can only submit the respective cells coordinates containing the actual  pick-up and drop-off locations. Fig.~\ref{fig1: map} illustrates an example of division of the state of Tennessee, USA into $27$ cells. Similarly, the exact pick-up/drop-off times can also be generalized (hidden) by using temporal cloaking where the actual time is generalized by setting \textit{time interval $T$}.
For instance, a driver $d$ who is interested in sharing a ride with others should do the following steps before sending his offer.
\begin{enumerate}
	\item Defines the planned trip $\Gamma^{(d)}$

$$\Gamma^{(d)}= \left\{\left(\ell_{0}^{(d)}, t_{0}^{(d)}\right),\ldots,\left(\ell_{k}^{(d)}, t_{k}^{(d)}\right), \ldots,\left(\ell_{n}^{(d)}, t_{n}^{(d)}\right)\right\}$$

	that consists of the exact departure location $\ell_{0}^{(d)}$, departure time $t_{0}^{(d)}$, destination $\ell_{n}^{(d)}$ and estimated arrival time $t_{n}^{(d)}$, as well as a sequence of optionally intermediate locations and their corresponding arrival times $\left(\ell_{k}^{(d)}, t_{k}^{(d)}\right)$.

	\item Then, hides or cloaks his exact trip locations and times to zones (cells) and time intervals as follows.
    
    \begin{equation}
        \Lambda^{(d)}=\left\{\left(C_{0}^{(d)},  T_{0}^{(d)}\right), \ldots, \left(C_{k}^{(d)},  T_{k}^{(d)}\right), \ldots, \left(C_{n}^{(d)}, T_{n}^{(d)}\right)\right\}
    \end{equation}
    
where $\left(C_{0}^{(d)}, T_{0}^{(d)}\right)$ represents the cloaked location and time that corresponds to $\left(\ell_{0}^{(d)}, t_{0}^{(d)}\right)$.
    
	\item Creates a set of all possible trips in his planned trip of the elements in $\Lambda^{(d)}$ as.
    \begin{equation}
    \label{driver table}
	\begin{aligned} \widehat{\Lambda^{(d)}}=\{\left(C_{j}^{(d)}, T_{j}^{(d)}, C_{k}^{(d)} \right)\} \\ &  1 \leqslant j \leqslant n  \\ & j+1\leqslant k \leqslant n    \end{aligned}
	\end{equation}
    
    where the number of all possible trips depends on the number of chosen points $n$ and it can be mathematically expressed as~\cite{BC}
  
    $$
\left( \begin{array}{l}{n} \\ {2}\end{array}\right)=\frac{n !}{2 !(n-2) !}
$$
\end{enumerate}

 For a rider $r$, we denote his ride request as $\Gamma^{(r)}=\left\{\left(\ell_{0}^{(r)}, t_{0}^{(r)}\right), \ell_{d}^{(r)}\right\}$, where $\left(\ell_{0}^{(r)}, t_{0}^{(r)}\right)$ represents the departure location  and the desired set off time,  and  $\ell_{d}^{(r)}$ denotes the drop-off location. Similar to the driver, the rider also converts the request into the generalized form by
mapping the pick-up and drop-off locations into the corresponding cells as well as departure time into cloaked time.

  \begin{equation}
  \label{7}
        \widehat{\Lambda^{(r)}}=\{\left(C_{o}^{(r)}, T_{o}^{(r)}, C_{d}^{(r)} \right)\}
    \end{equation}


Note that in all the previous steps, including cloaking of the drivers' trip (see Table 4) and the rider trip, are done off the blockchain using, for instance, the driver/rider smart phones.


\begingroup\renewcommand{\caption}[1]{}%
			\begin{figure*}[!t]
		\centering
		\includegraphics[width=.9\linewidth]{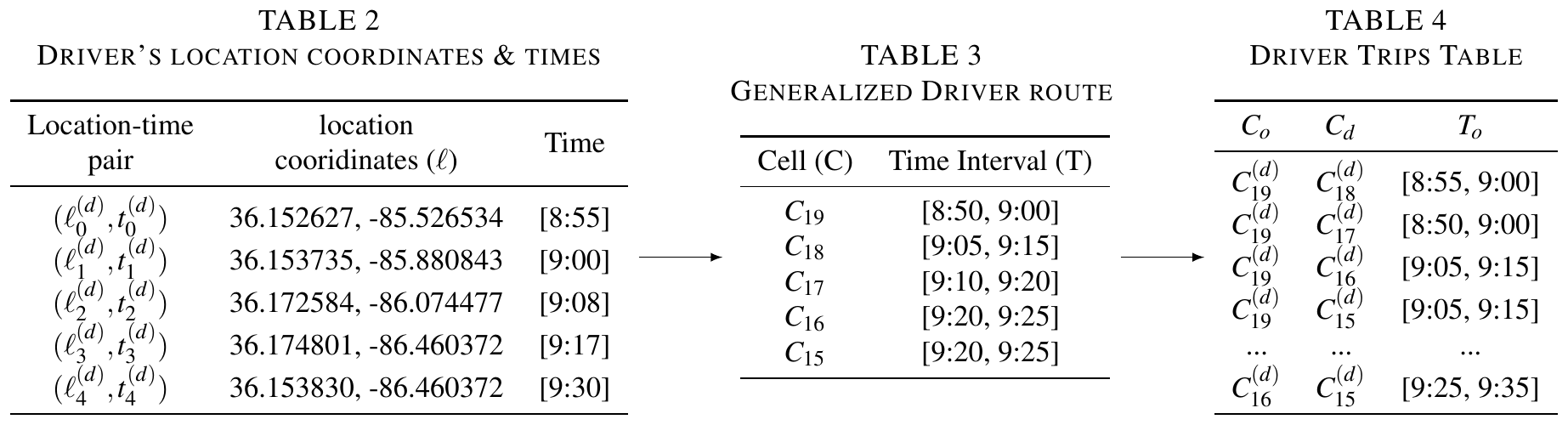}
		\vspace{0mm}
		
		\caption{Illustration of the bidding and selection phase in B-Ride. }
	\end{figure*}
\endgroup

\subsection{Bidding and Selection}
In this section, we describe the process of matching riders' requests with drivers' offers atop public blockchain. The smart-contract $\mathcal{L}$ that  handles the B-Ride logic including bidding and selection is summarized in Algorithm \ref{alg1}. The different steps required for bidding and selection are illustrated by a schematic diagram in Fig.~\ref{Bidding}, and detailed in the following sections.

\subsubsection{Publishing the ride request}
First, as a fundamental concept to avoid de-anonymization in the blockchain, every rider $r$  uses for each ride request a new blockchain address $\mathsf{ADD}^{(r)}$ that corresponds to a fresh  public/private key pair ($PK^{(r)}, SK^{(r)}$).
Then, a rider publishes a ride request that contains his/her cloaked pick-up/drop-off $(C_{o}^{(r)},  C_{d}^{(r)})$ and cloaked time $T_{o}^{(r)}$. Also, the request should include an expiry time to receive driver's offers. Optionally, the request can include a maximum number of offers to be received. Note that this request should be signed by the temporary private key of the rider and sent to the smart-contract by making a call to the function \textit{MakeRideRequest()} in Algorithm~\ref{alg1}. Once the miners validate the corresponding signature of the rider, the request will be public to all drivers.


\subsubsection{Submitting drivers' offers}
For a driver that wants to share a ride, first needs to first receive a public key certificate from the corresponding registration authority (RA), such as the government. A driver $d$ having a unique identity (e.g., license plate number), creates a public-secret key pair ($PK^{(d)}, SK^{(d)}$) and registers at the RA to obtain a certificate binding $PK^{(d)}$ to $d$. 

 Drivers can either periodically query the blockchain to obtain new riders' requests or use some out-of-band signaling protocols to get notified each time a new  request is published~\cite{knirsch2018privacy}. In order to make an offer, a driver $d$  first verifies if the spatio-temporal attributes of a received request made by a rider $r$ falls within one of his own planned trips, (i.e., $\widehat{\Lambda^{(r)}} \in \widehat{\Lambda^{(d_i)}}$).  
 If one of the requests matches the driver's trip, as shown in Fig.~\ref{Bidding}, the driver creates an offer that should include all the necessary information for the rider such as, the exact pick-up location and time $\left(\ell_{0}^{(d_i)}, t_{0}^{(d_i)}\right)$, the exact drop-off location and time  $\left(\ell_{d}^{(d_i)}, t_{d}^{(d_i)}\right)$ as well as the offer bid price $\mathcal{B}_{d_i}$ (e.g., price per mileage). Then, the driver uses the rider's public key to encrypt all above information to obtain $\mathcal{C}_{d_i}$

$$\mathcal{C}_{d_i}= \mathcal{E}_{PK^{(r)}}\left(\ell_{0}^{(d_i)}, t_{0}^{(d_i)}, \ell_{d}^{(d_i)}, t_{d}^{(d_i)}               \right) $$

Where $\mathcal{E}$ is an asymmetric encryption algorithm e.g., RSA. This is necessary to preserve the privacy of both the driver and rider.
The tuple $\mathcal{C}_{d_i}\|\mathcal{B}_{d_i}$ is sent by the driver to the smart-contract by calling the function \textit{MakeRideOffer()}. Note that the bidding price is not encrypted and made  public in order to ensure price competition which guarantees low prices for the riders. Also that since the rider request includes an expiry time to receive offers, any offer made after that time will be automatically rejected by the blockchain network. 

\subsubsection{Finding Feasible Matches}
 After receiving, for instance, $n$ offers $\{\mathcal{C}_{d_1},\cdots,\mathcal{C}_{d_n}\}$ for the same request made by a rider $r$, the rider retrieves the encrypted offers from the smart-contract and decrypts each of them off the chain using his secret key. In order to select an offer that better matches the rider preferences, the offers are evaluated as follows.
\begin{enumerate}
    \item The driver's pick-up and drop-off  match (i) \textit{spatially} by checking if:

          \begin{equation}
\left(\abs {\ell^{(d_i)}_{o}-\ell^{(r)}_{o}} \leq \delta^{(r)} \right)\land \left(\abs{\ell^{(d_i)}_{d}-\ell^{(r)}_{d}} \leq \delta^{(r)}\right),              
          \end{equation}
where $\delta^{(r)}$ is the maximum distance that the rider can  walk to meet the driver's pick-up location, or to reach his final destination.

           (ii) \textit{Temporarily} by checking if:

          \begin{equation}
\left(\abs{t^{(d_i)}_{o} - t^{(r)}_{o}} \leq \tau^{(r)}\right) \land \left(\abs{t^{(d_i)}_{d} - t^{(r)}_{d}}\leq \tau^{(r)}\right),           
          \end{equation}
 where  $\tau^{(r)}$ is the maximum delay the rider can tolerate before meeting the driver at the pick-up location, or to reach his final destination after being dropped-off.
   
  \item Besides, in B-Ride, each driver $d$ has a reputation value $\beta_{d_i}$ that is stored on the blockchain. The detail about how each driver receive his reputation will be  discussed in details in the next sections.
  
 Using $\delta^{(r)}$, $\tau^{(r)}$, $\mathcal{B}_{d_i}$ and the drivers' reputation score, the rider is able to select the best offer that match his preferences as follows: 
\[\argmin_{\forall d_i } (\delta^{(r)}, \tau^{(r)}, \mathcal{B}_{d_i}) \land  \argmax_{\forall d } (\beta_{d_i})\]

Note that preferences may vary from one rider to others. For instance, some may prefer an offer with a low price even if it has a high space slack $\delta^{(r)}$ or waiting time $\delta^{(r)}$. Different from existing centralized approaches, finding feasible ride matches is handled over the blockchain and is therefore, fully transparent.

\end{enumerate}


\subsubsection{Illustrative example for the selection/matching phase}

To better illustrate the selection process, let us assume a driver $d_i$ who has a trip as shown in Fig~\ref{fig1: map}. The driver's route  starts from Knoxville and ends in Nashville with four stop points that lies on his route. Note that defining these points depends only on the driver preferences. Table ~2 illustrates the coordinates of the stop points with their corresponding arrival times which are defined by the driver. As discussed previously in Sec.~\ref{trips}, the driver's route should be generalized/mapped as given in Table~3 with respect to cell division of Tennessee given in Fig.~\ref{fig1: map}. Thus, the driver creates his own trip table that contains possible trips that he/she is willing to share with other riders, as indicated in Table~4. 

On the other hand, a rider $r$ who is looking to take a ride with the following attributes:
$$\Gamma^{(d)}= (35.222;-100.1511, 4548754, 35.221;-100.1511)$$
publishes his request to the blockchain by sending a generalization version of his ride:
        $$\widehat{\Lambda^{(r)}}=\{\left(Utica, 45222211, Syracuse\right)\}$$
 
Each driver, therefore, reads the existing requests on the blockchain to decide the offers they could submit. For instance, based on the trip table of the driver $d_i$ in Table 4, he/she can determine that the rider's request is in its route i.e., the 4th element in Table~4. Thus, he/she follows up on this request by sending an offer that contains the exact pick-up, drop-off and time encrypted to the rider. Finally, the rider evaluates the offer and compares it with any other offers to select the best matched one.

\begin{algorithm}[!t]
 \algorithmfootnote{Note that for the sake of
 explanation, we separated Algorithm~\ref{alg:contract2} and Algorithm~\ref{alg:paymentcontract}. In practice, B-Ride can be entirely implemented in a single smart-contract.}
\SetKwProg{Fn}{function}{}{}

\SetKwProg{Contract}{contract}{}{}
\SetKwData{NumOfUpdatedObjects}{numOfUpdatedObjects}
\SetKwIF{If}{ElseIf}{Else}{if}{}{else if}{else}{end if}
\SetKwFunction{PaymentContract}{PaymentContract}
\SetKwFunction{RecieveRideRequest}{MakeRideRequest}
\SetKwFunction{RecieveRideOffer}{MakeRideOffer}
\SetKwFunction{BRide}{BRide}
\SetKwFunction{TimeLockedDeposit}{TimeLockedDeposit}
\SetKwFunction{BiddingSelectionContract}{BiddingSelectionContract}
\Contract{\text{BRide}}{
\textcolor{blue}{mapping}{(\textcolor{blue}{address} => int) Reputation score\_1}
\tcp{Mapping for drivers' reputation score of arriving to pick-up}

\textcolor{blue}{mapping}{(\textcolor{blue}{address} => int) Reputation score\_2}
\tcp{Mapping for drivers' reputation score of completed trips}
  \BlankLine
\Contract{\BiddingSelectionContract}{
  \Fn{\RecieveRideRequest{$C^{(r)}_0$,$C^{(r)}_d$,$T^{(r)}_0$}}{
  \tcp{Receive ride request.}
  }
   \Fn{\RecieveRideOffer{$\mathcal{C}^{(d_{i})}$,$\mathcal{B}_{d_{i}}$}}{
  
  \tcp{Receive ride drivers biddings.}
 } }
  \BlankLine
\Contract{\TimeLockedDeposit}{
Create a sub-contract of Algorithm~\ref{alg:contract2}
}
  \BlankLine   
\Contract{\PaymentContract}{
Create a sub-contract of Algorithm~\ref{alg:paymentcontract}
} }
\caption{Pseudocode for \textit{B-Ride} contract}
\label{alg1}
\end{algorithm}

\subsection{Time-locked Deposit Protocol}
In this section, we introduce the time-locked deposit protocol that prevents malicious drivers or riders from not committing to their respective ride offers or requests. The deposit serves as a guarantee for their good intention.
 Usually, a traditional solution to this problem  is to allow both parties (i.e., drivers and riders) to pay a subscription fee to a trusted agency that can be contacted  whenever a breach occurs. However, this solution may fail since an honest party has to expend extra effort by contacting and convincing the trusted party about the breach ~\cite{bentov2014use}. In addition, if the third party gets attacked, the whole system becomes unprotected against malicious users. 
 
\begin{figure}[!t]
		\centering
		\includegraphics[width=.9\linewidth]{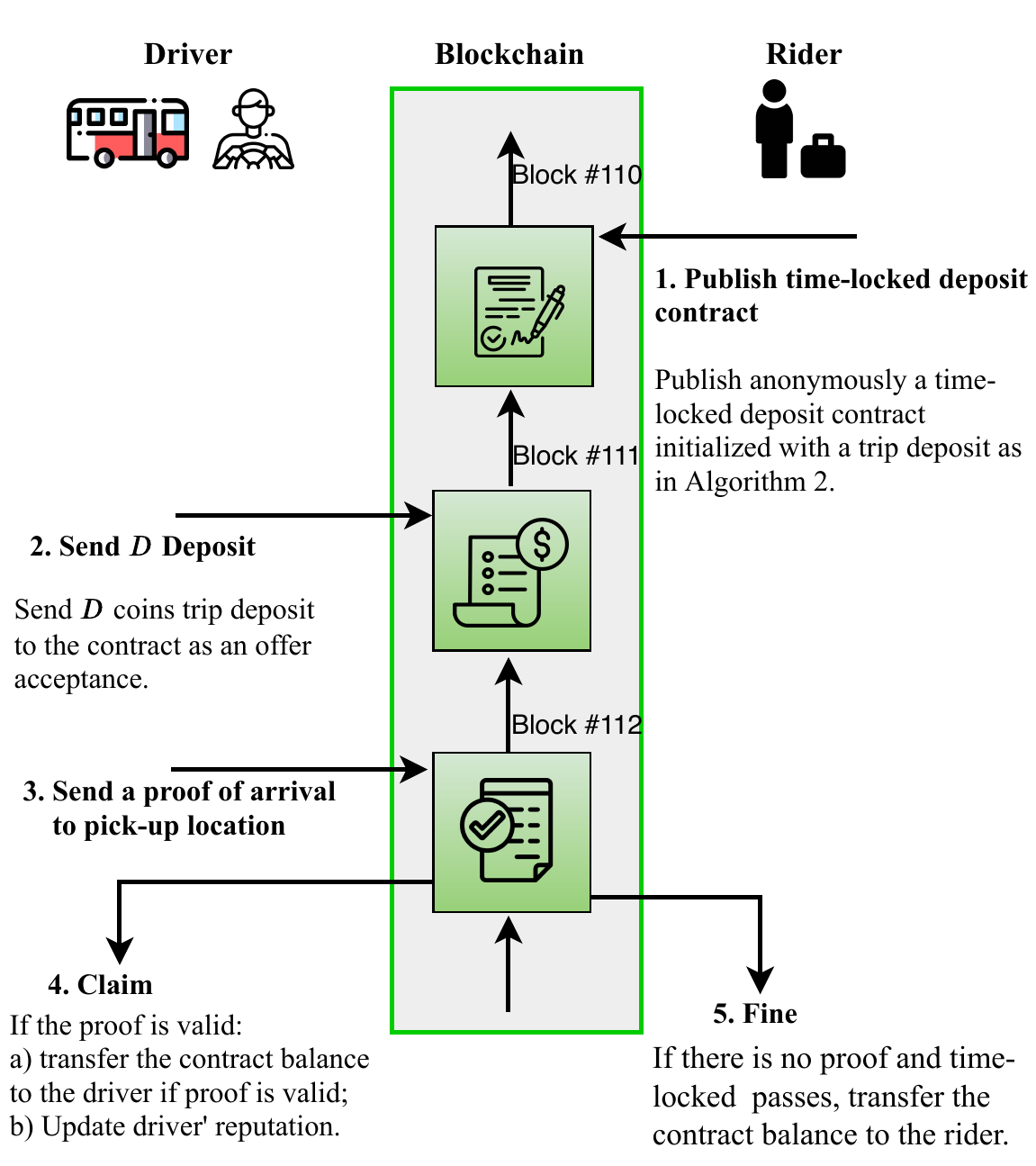}
		\vspace{0mm}
		\caption{The schematic diagram of B-Ride the protocol as proof-of-concept.}
		\label{timelocke}
\end{figure}

Inspired by~\cite{bentov2014use}, we propose a time-locked deposit protocol for ride-sharing service leveraging smart-contracts and blockchain. The smart-contract receives a deposit from both rider and driver and  conditionally transfers the total deposit amount to the "driver" if he/she arrives to the rider pick-up location at the predefined agreed time. However, if the driver defaults, the deposit will be  transferred to the rider after a prespecified time as a fine to the driver. Note that the conditions are defined in the smart-contract and executed over the blockchain, in a completely transparent and secure manner. Nevertheless, the definition of these conditions should consider the privacy of the drivers and riders. In other words, how a driver can prove that he/she has arrived to the  rider's pick-up location (so the blockchain enforces rewards) without revealing such sensitive information on the public blockchain. For this purpose and in order to preserve users' privacy, we leverage the use of the zero-knowledge set membership proof (ZKSM) protocol presented in Sec.~\ref{ZKSM}. In nutshell, after selecting an offer from a driver, the rider publishes a new \textit{time-locked deposit} smart-contract (see Algorithm.~\ref{alg2}), and initializes it by sending the set of locations  as well as the trip deposit. The corresponding driver of the selected offer should act as a \textit{prover} who needs to demonstrate to the blockchain (\textit{verifier}) in a \textit{zero-knowledge} manner that he/she has arrived to one of the pick-up locations already published by the rider  without revealing the exact pick-up location. A schematic diagram of the time-locked deposit protocol is depicted in Fig.~\ref{timelocke}. 
 
In the following we describe the different steps of the time-locked deposit protocol.

 \subsubsection{Initialization}
The rider initializes the time-locked deposit protocol as follow:
\begin{enumerate}
    \item Defines a set $\phi$ of $k$ locations where $\phi=\{\ell_1,\cdots,\ell_k\}$. The set $\phi$ should include the actual rider pick-up location $\ell^{(r)}_{o}$, as well as other obfuscated locations. 
    \item Picks a random number $x \in_{R} \mathbb{Z}_{p}$ and computes a corresponding public $y \in g^{x}$, where $g$ is the generator of the order $-q$ subgroup of $\mathbb{Z}_{p}$
    \item Computes for every element $i\in \phi$ the corresponding signature using $g$ and $x$: where, $\mathcal{A}_i = g^{\frac{1}{(x+i)}}$.
    \item Publishes a new smart-contract, given in Algorithm~\ref{alg2}, and calls the function \textit{TimeLockedDeposit($\phi$, $A_i$, deposit)}.
\end{enumerate}

\begin{figure*}[!t]
		\centering
		\includegraphics[width=1\linewidth]{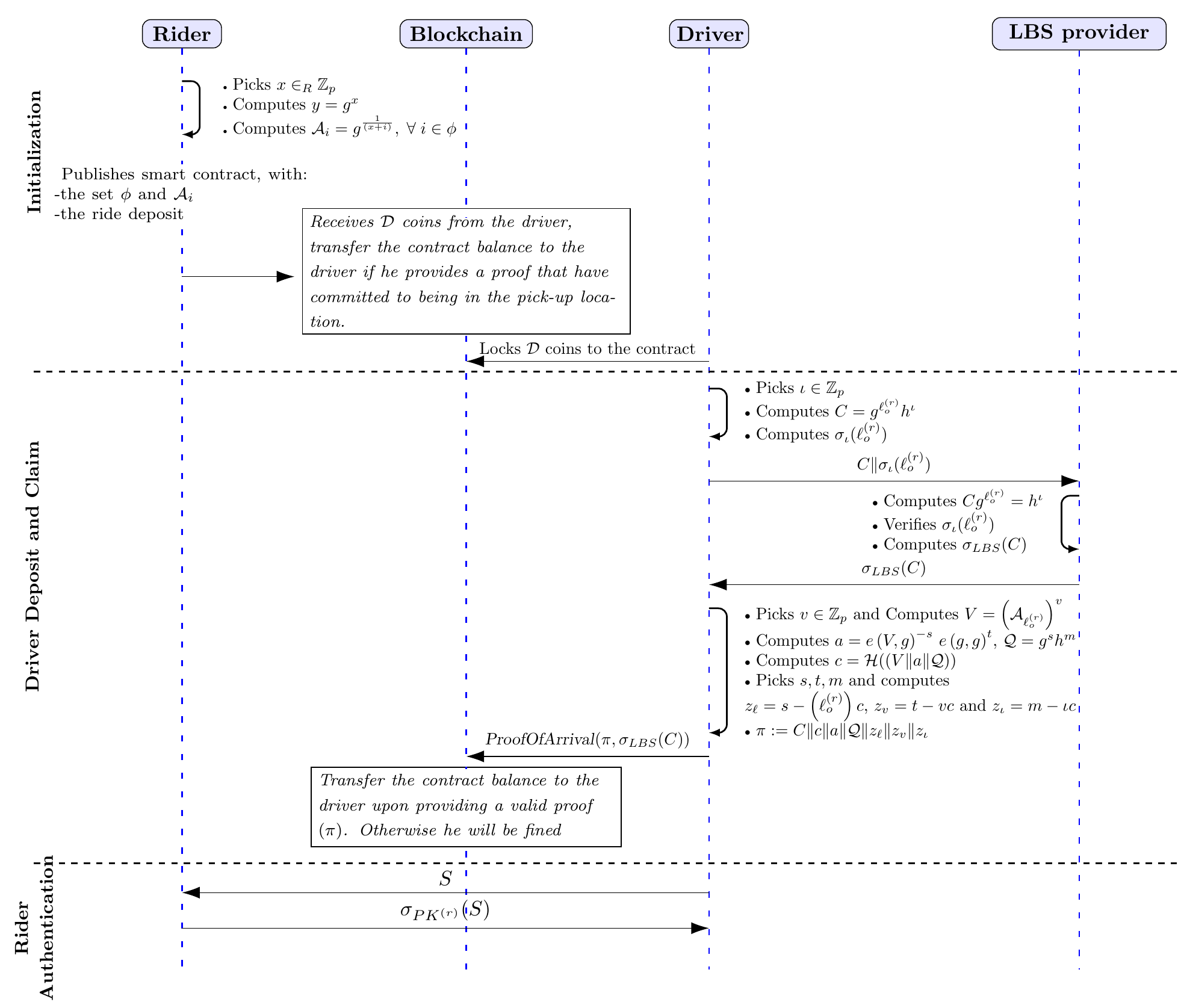}
		\vspace{0mm}
		
		\caption{Time-locked deposit protocol in B-Ride. }
		\label{3}
\end{figure*}

\subsubsection{Driver Deposit and Claim}
In this step, the driver needs to confirm his/her commitment to the trip offer by  sending a deposit to the smart-contract. The driver can claim back the deposit, once he/she arrives to the pick-up location, by providing a proof of arrival to that location at the pick-up time. 

For this purpose, before sending the deposit, the driver first authenticates each element in $\phi$ using the corresponding rider's public key\footnote{The purpose of this verification is to prevent a  malicious rider from falsely setting-up the ZKSM protocol so he/she cannot manipulate the rewards of the driver.}. The verification is done by checking the following equality~\cite{boneh2004short}
 $$e\left(\mathcal{A}_i, y \cdot g^{\ell_{i}}\right)\stackrel{?}{=}e(g, g)$$
To send the deposit, the driver calls the function \textit{DriverDeposit($\mathcal{D}$)}, where $\mathcal{D}$ is a deposit to the contract within a limited window of time (See line 15 in Algorithm \ref{alg2}). 

When the driver reaches the pick-up location, he/she needs to prove in zero-knowledge that he/she arrived to the pick-up location as follow:
  \begin{enumerate}
        \item  The driver picks $v \in_{R} \mathbb{Z}_{p}$ and computes $V = \left(\mathcal{A}_{\ell^{(r)}_{o}}\right)^v$, where $A_{\ell^{(r)}_{o}}$ is the rider' signature on $\ell^{(r)}_{o}$. Then, he/she computes a commitment on $\ell^{(r)}_{o}$  using Pedersen commitment~\cite{pedersen1991non} as $C=g^{\ell^{(r)}_{o}}h^{\iota}$ where $\iota$ is a random number and $h$ is a random group element such that it is hard to find the discrete logarithm of $g$ base $h$. The tuple $C \| \ell^{(r)}_{o} \| \sigma_{\iota}(\ell^{(r)}_o)$ is then sent to the the location prover (LP)  where $\sigma_{\iota}(\ell^{(r)}_o)$ is the driver' signature on $\ell^{(r)}_o$.
  
        \item The LP verifies if the received location from the driver is authentic. The LP Computes $$C'=Cg^{-\ell^{(r)}_{o}}=g^{\ell^{(r)}_{o}}h^{\iota}g^{-\ell^{(r)}_{o}}=h^{\iota}$$ where the $h^{\iota}$ is the public key corresponding to $\iota$. Then, the LP verifies if the signature $\sigma_{\iota}(\ell^{(r)}_o)$ is valid using $h^{\iota}$. The tuple $C \| \sigma_{LP}(C)$ is then sent back to the driver, where $\sigma_{LP}(C)$ is the LP signature on $C$.
  
        \item The driver picks random numbers $s, t, m \in_{R} \mathbb{Z}_{p} $ and computes
         $$a= e\left(V,g\right)^{-s} e\left(g,g\right)^{t}, \text{and}\; \mathcal{Q}=g^{s}h^{m}$$
        Note that the adopted ZKSM in~\cite{camenisch2008efficient} requires an interactive session between the prover and the verifier with multiple rounds of communication. However, in the context of blockchain, miners, can not properly agree on a common value of  proof-related parameters, such as the challenges, since they need to be chosen randomly. One solution is to employ the FiatShamir heuristic~\cite{fiat1986prove}, which is a generic technique that allows to convert interactive zero-knowledge schemes to non-interactive protocols. Using the Fiat-Shamir heuristic, a challenge $c$ can be calculated from public parameters as follows
$$c=\mathcal{H}(V\|a\|\mathcal{Q})$$

Then, he/she computes $z_{\ell}=s-  \left(\ell^{(r)}_{o}\right) c$, $z_{v}=t- v c$ and $z_{\iota}=m- \iota c$. 

Then, the proof of ZKSM is denoted as $\pi= C\|c\|a\|\mathcal{Q}\|z_{\ell}\| z_{v}\| z_{\iota}$. Finally, he/she sends $\pi\|\sigma_{LP}(C)$ as a transaction to the blockchain.
         
    \item Once the contract $\mathcal{T}$ receives the proof $\pi$, it should verify whether $(i)$ the proof is from the selected driver, and $(ii)$ the time of receiving the proof lies on the generalized rider pick-up time (See lines 19-20 in Algorithm \ref{alg2}). 
   The proof that $C$ is a commitment to an element in $\phi$ can be validated by the following statement~\cite{camenisch2008efficient}.
   \begin{equation}
   \label{ZK}
\operatorname{\textbf{PK}}\left\{(\ell^{(r)}_{o}, \iota , v) : C=g^{\ell^{(r)}_{o}} h^{\iota} \wedge V=g^{\frac{v}{x+\ell^{(r)}_{o}}}\right\}       
   \end{equation}

Proving the statement in~(\ref{ZK}) is done in the blockchain by checking the following conditions: 
          \begin{equation}
              \mathcal{Q}\stackrel{?}{=} C^{c}h^{z_{\iota}}g^{z_\ell}
              \label{c1}
          \end{equation}
          \begin{equation}
              a\stackrel{?}{=} e\left(V,y\right)^{c}\cdot e\left(V,g\right)^{-z_{\ell}} \cdot e\left(g,g\right)^{z_{v}}
              \label{c2}
          \end{equation}

Once the function $\textit{ProofOfArrival}$ validates the two conditions in \ref{c1} and \ref{c2}, the driver claims the contract balance (see lines 20-26 in Algorithm~\ref{alg2}). Due to the \textit{Soundness} and \textit{Completeness} property of ZKSM~\cite{camenisch2008efficient}, the contract accepts the proof if it is correctly constructed by the driver. If the driver breaks his/her commitment to arrive in the pick-up time, which is pre-defined in the time-locked deposit contract, he/she will be fined with a monetary penalty that goes to the rider' account (see lines 27-30 in Algorithm~\ref{alg2}). 
  \end{enumerate}

  \subsubsection{Driver/rider authentication}
  In order to prevent an impersonation attack, in which a malicious rider tries to take a ride that was reserved by another rider, driver and rider must authenticate each other. Specifically, the rider should prove to the driver in zero knowledge that he/she indeed knows the value of private key corresponding to the public key ($PK^{(r)}$) that made the reservation. The driver (the verifier) selects a uniformly random integer $\mathcal{S} \in \mathbb{Z}_{p}$ as a \textit{challenge} and sends it to the rider (the prover). The rider uses his/her private key to generate a signature on the challenge ($\sigma_{SK^{(r)}}(\mathcal{S})$) and sends $(\mathcal{S}\|\sigma_{SK^{(r)}}(\mathcal{S}))$ to the driver. Finally, the driver verifies if $VerifySig(PK^{(r)},\sigma_{SK^{(r)}}(\mathcal{S}),\mathcal{S}) = 1$. 
	
   \begin{algorithm}[!t]
\algsetup{linenosize=\small}
 \scriptsize
\SetKwProg{Fn}{function}{}{}
\SetKwProg{Contract}{contract}{}{}
\SetKwData{NumOfUpdatedObjects}{numOfUpdatedObjects}
\SetKwIF{If}{ElseIf}{Else}{if}{}{else if}{else}{end if}
\SetKwFunction{FirmwareUpdateContract}{FirmwareUpdateContract}
\SetKwFunction{TimeLockedDeposit}{TimeLockedDeposit}
\SetKwFunction{ProofOfArrival}{ProofOfArrival}
\SetKwFunction{DriverDeposit}{DriverDeposit}
\SetKwFunction{FineDriver}{FineDriver}
\Contract{\TimeLockedDeposit}{
\textcolor{blue}{uint} public Balance \tcp{Balance to withhold driver and rider deposits}
  \textcolor{blue}{address} payable rider \tcp{Rider address}
      \textcolor{blue}{address} payable driver \tcp{driver address}
      \textcolor{blue}{uint} public RiderDeposit  \tcp{RiderDeposit}
      \textcolor{blue}{uint} public DriverDeposit \tcp{DriverDeposit}
      \textcolor{blue}{address} [] Set \tcp{set of obfuscated locations}
     \textcolor{blue}{address} [] $A_i$ \tcp{Signatures of elements in the set}
\tcp{Constructor}
  
  \Fn{\TimeLockedDeposit{\_driver, \_Set, $\_A_i$, \_RiderDeposit}}{
          driver $\leftarrow$ \_driver; \tcp{The address of selected driver}
        Balance  $\leftarrow$ \_RiderDeposit ;\tcp{This deposit acts as acceptance to the driver offer.}
             Set $\leftarrow$ \_Set;  \\
              $A_i$ $\leftarrow$ $\_A_i;$ 
  }
  \BlankLine
  
    \Fn{\DriverDeposit{uint256 \_DriverDeposit}}{
     \tcp{Receive Driver Deposit and add to the contract}
     \lIf {block.timestamp $\geq$ expiration}{return}
     
     \lIf {msg.sender $\neq$ DriverAddress}{return}
        \lIf {msg.value $\neq$ DriverDeposit}{return}
          \lIf {now $\geq$ $T^{Accept}_{deadline}$}{return}  \tcp{$T^{Accept}_{deadline}$ is a window for the driver to send his/her deposit e.g., 2 min}
       Balance $\leftarrow$\_DriverDeposit;
  } 
  \Fn{\ProofOfArrival{($\pi$, $\sigma_{LP}(C)$]}}{
 \tcp{$\pi=\{D\| c \| a \| \mathcal{Q}\| z_\sigma\| z_v\| z_r\}$}
  \tcp{If the driver provides a valid proof of the agreed pickup location, he/she will be rewarded by the rider deposit and he/she get his/her deposit back.}
\lIf {msg.sender $\neq$ DriverAddress}{return}
\lIf {now $\neq$ $T^{(r)}_{o}$}{return} \tcp{Check the time of receiving the proof lies in the generalized rider pick-up time}

   \If{(ZKSM.Verifier($\pi$))}{
   \tcp{ZKSM.Verifier is a library embedded in the runtime
environment of smart contract such as EVM}

           BRide.Reputation score\_1[msg.sender]$\leftarrow $ Reputation score\_1[msg.sender]+1;
           \tcp{Increase the driver reputation score of arriving to pick-up location}
           
           \tcp{call public event \textcolor{red}{Transfer} to finalize the payment}
           transfer(balance, driver);
            }
  }
  \BlankLine
  
  \Fn{\FineDriver{}}{
  
  \tcp{Issue a rider deposit back and the driver deposit and if the timeout has expired.}
  
   \lIf {block.timestamp $<$ expiration}{return}
      \lIf {msg.sender $\neq$ rider}{return}
      transfer(balance, rider); \tcp{Transfer the contract balance  back to the rider account.}
  
  } 
}
\caption{Pseudocode for \textit{time-locked deposit} contract $\mathcal{T}$ in B-Ride}\label{alg:contract2}
\label{alg2}

\end{algorithm}

\subsection{Fair Payment in Trust-less Environment}

In this section we address the following challenge: \textit{how to estimate the fare of each trip without trusting riders and drivers?} 

In current RSSs, to estimate the trip fare, a driver needs to report the distance and duration of each ride using his/her personal smartphone to the service provider. However, smartphones are general-purpose devices and are easy to tamper with. Thus, a malicious driver can report longer ride distances to get a higher fare. This problem is usually referred to as \textit{overcharging}~\cite{pham2017oride},~\cite{over}.

In B-Ride, we tackle the above challenge by adopting the \textit{pay-as-you-drive} method. After starting the trip, the rider makes a call to the \textit{fare payment}  smart-contract (see Algorithm \ref{alg:paymentcontract}) and initializes it with a sufficient amount of coins as a deposit to be used later for the payment of the fare. Note that the down payment used in the time locked deposit protocol is also used as a part of the payment of trip's fare. After that, for every period of time, the driver sends the elapsed distance to the rider. The rider checks whether the received distance matches the actual elapsed distance or not. Using multi-signature, the elapsed distance is signed by both the driver and the rider, and then sent to the smart contract as a proof of the actual elapsed distance approved by the driver and rider. Thus, a payment that corresponds to that travelled distance, will go to the driver account. This ensures that the driver is paid based on the distance that has been actually travelled. In case a malicious rider stops sending the proofs of the elapsed distance, the driver can decide to stop the ride without affecting his/her payment on the previous travelled distances. Similarly, if the driver decides to stop the trip, the rider can stop sending the elapsed distance proof and, therefore, the driver will not be paid. Finally, if the driver does not complete the trip in the a pre-agreed time of the driver's offer, the remaining payment can go back to the rider's account (See function Refund in algorithm~\ref{alg:paymentcontract}). This enforces the driver to commit to his/her offer and complete the journey on time without delay. 

\begin{algorithm}[!t]
\algsetup{linenosize=\small}
 \scriptsize
\SetKwProg{Fn}{function}{}{}
\SetKwProg{Contract}{contract}{}{}
\SetKwData{NumOfUpdatedObjects}{numOfUpdatedObjects}
\SetKwIF{If}{ElseIf}{Else}{if}{}{else if}{else}{end if}
\SetKwFunction{FirmwareUpdateContract}{FirmwareUpdateContract}
\SetKwFunction{RidePayment}{RidePayment}
\SetKwFunction{ProofOfDistance}{ProofOfDistance}
\SetKwFunction{withdrawFunds}{withdrawFunds}
\Contract{\RidePayment}{

  \textcolor{blue}{address} payable rider \tcp{Rider address}
      \textcolor{blue}{address} payable driver \tcp{driver address}
      \textcolor{blue}{uint} public TripDist \tcp{driver address}
\tcp{Constructor}
  \BlankLine
  \Fn{\RidePayment{\_driver,  \_TripDist, \_$t^{(R)}_d$}}{
    \_driver $\leftarrow$ \_driver \tcp{A greed distance of payment} 
         TripDist $\leftarrow$ \_TripDist \tcp{A greed distance of payment} 
  }

  \BlankLine
    
  \Fn{\ProofOfDistance{ElapsedDist}}{
  
  \tcp{If this is called with the rider, the driver gets paid.}
  
  \lIf {msg.sender $\neq$ RiderAddress}{return}
    \While{TripDist $\leq$ ElapsedDist }{
    
   transfer(balance $\times$  ElapsedDist), pk\_d) \tcp{Decrease balance according to the travelled distance by the rider and driver}
    
     TripDist $\leftarrow$ \_TripDist-Elapsed\_Distance \;

        \If{(TripDist==0)}{
           B-Ride.Reputation score\_2[DriverAddress]$\leftarrow $ Reputation score\_2[DriverAddress]+1;\\
           \tcp{Update the reputation driver score of completed trips}
            }
    }
  }
 \BlankLine
  
  \Fn{\withdrawFunds{}}{
  
  \tcp{Issue a refund back to the rider if the timeout has expired.}
  
   \lIf {block.timestamp $<$ expiration}{return}
      \lIf {msg.sender $\neq$ owner}{return}
      transfer(balance, owner) \tcp{Transfer the contract balance  back to the rider account.}
  
  }
  }
\caption{Pseudocode for the \textit{fair payment} contract in B-Ride}\label{alg:paymentcontract}
\end{algorithm}

\subsection{Reputation Management}
In B-Ride, before starting the trip, the driver needs to prove his/her arrival to the pickup location, and the rider is required to prove his/her good willing by making a down-payment. However, in case the ride was not performed, the system cannot determine which side is circumventing (not committed to the trip), whether the rider or driver. In addition, because the down-payment made by a rider automatically goes to the driver once he/she proves his/her arrival to the pick-up location, a dishonest driver can collect the down-payments without committing to his/her offer. Also, since the riders are anonymous, it is not possible to make a claim about such situation. To mitigate this problem, we propose a reputation system that records any similar incidents as a potential driver misbehavior. If for the same driver this case happens many times, the reputation score of the driver reduces and this will discourage riders from accepting any offers coming from that driver.

More specifically, each driver is assigned a reputation score that will be considered by riders during the selection process. A high reputation score reflects the good driver behaviour. Moreover, unlike traditional ride sharing schemes where the reputation is managed and controlled by a third party, our reputation system is completely decentralized and implemented on blockchain.
Each driver is tagged with a reputation score $\beta_d$. The score has two values, the first value $\beta^{AP}_D$ increments every time the driver prove his/her arrival to the agreed pick up location, and the second one $\beta^{AD}_d$ increases for every received proof of completing the trip. 

 The driver's reputation $\beta^{AP}_d$ increases and is recorded in blockchain if the function \texttt{ProofofArrival} validates his/her proof of arriving to the rider pick up location (See line 24 in Algorithm~\ref{alg:contract2}). Meanwhile, the driver's reputation score $\beta^{AD}_{d}$ increases only if the trip is completed. (See line 14 in algortihm~\ref{alg:paymentcontract}). Therefore, when a dishonest driver $d$ tries to collect riders deposits without completing the corresponding trips, the $\beta^{AP}_{d}$ increases while $\beta^{AD}_{d}$ keeps decreasing and becomes smaller than $\beta^{AD}_{d}$. Thus, the final reputation score $\beta_{d}$ of driver $d$ can be estimated as follow:
 
 \begin{equation}
\beta_{d} = \dfrac{\beta^{AP}_{d}}{\beta^{AD}_{d}}
\end{equation}

Note that it is possible for a new driver joining the system, that does not have a reputation score yet, to be selected by some riders. In case this driver may be dishonest by collecting some down-payment before his/her reputation score get decreased. To limit this to happen, each newly registered driver needs to put a deposit that will be used in the future to refund any riders that lost their down-payments for the account of that driver, in case the driver will receive a bad reputation score. 

The value of the reputation score allows to distinguish between three different cases: $(i)$ $\beta_d  =  1$ means that $\beta^{AP}_{d} = \beta^{AD}_{d}$. In this case, the driver can be considered honest as he/she has committed to all his/her past trips; $(ii)$ $\beta_d  \leq  \mathcal{T}$, where $\mathcal{T}$ is a threshold value pre-defined by the system users i.e., the riders. In this case, the driver can be considered dishonest and the system will refund all riders from whom he/she took the down-payment deposit, and $(iii)$ $ \mathcal{T}<\beta_d <1$. In this case, the driver is still considered honest but riders may prefer to select drivers with better score for their future trips.
   
 \begin{figure*}
      \centering
\begin{tikzpicture}
\node (tb1) {
	\begin{tabular}{m{3.1cm}||m{2.3cm}m{2.3cm}m{2.2cm}m{2.3cm}m{2.3cm}}
		\hline
		& Architecture & Rider's privacy& Trust & Fair payment & Transparency \\ 
		\hline\hline 	
			Current RSS~\footnote{https://www.blablacar.com/}& Centralized & $\times$ & $\bigcirc$ & $\times$ & $\times$\\\hline
		\hline
		SRide~\cite{aivodji2018sride} & Centralized & $\surd$  & $\times$ & $\times$ & $\times$ \\\hline
	Co-utile~\cite{sanchez2016co}	& Decentralized & $\surd$  & $\times$ & $\times$ & $\times$\\\hline
		DACSEE~\cite{DACSEE}, Arcade City~\cite{ArcadeCity}& Blockchain & $\times$ & $\times$ & $\times$ & $\surd$ \\\hline
			B-Ride	& Blockchain & $\surd$  & $\surd$   & $\surd$  & $\surd$  \\\hline
\end{tabular}};

\node[above=1.5ex,align=center] at (tb1.north) {TABLE 5: Comparison between our system B-Ride and existing RSS platforms.};

\node[below=0.2ex,align=left] at (tb1.south) {Note: $\surd$ denotes a realized functionality; $\bigcirc$ denotes a (partially) realized function by relying on a central trust\\ and  $\times$ denotes an unrealized feature.};
\end{tikzpicture}
\end{figure*}

\section{Performance Evaluations}
\label{comm overhead}
In this section, we evaluate the performance of B-Ride. We first assess the  functionalities of B-Ride by comparing it with existing riding systems. Then, we present a proof-of-concept implementation of B-Ride and demonstrate its feasibility. Finally we  evaluate the cost of the proposed implementation in term of computation and storage overhead. 
	
\begin{figure}[!t]
\centering
    \includegraphics[width=1\linewidth]{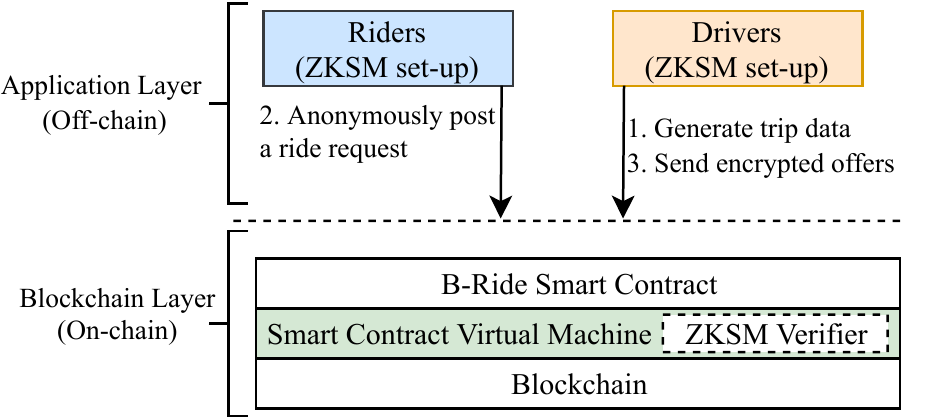}
    \vspace{0mm}
\caption{Implementation overview of B-Ride.}
\label{Fig. implment}
\end{figure}

\subsection{Functionality}
In Table~5, we summarize the existing ride sharing systems in terms of architectures and desired functionalities, such as riders privacy, trust, payment fairness and transparency. In~\cite{sanchez2016co}, a decentralized ride sharing system is proposed using peer to peer communications. However, the scheme is vulnerable to DDoS and Sybil attacks and more importantly lacks transparency that is achieved in B-Ride. Some projects already start the development of  blockchain-based ride sharing platforms e.g., DACSEE~\footnote{https://dacsee.com/} arcade city~\footnote{https://arcade.city/}, however, none of them  consider riders' location privacy or payment fairness.   

\subsection{Implementation}
Fig.~\ref{Fig. implment} shows an overview of the proof-of-concept implementation of B-Ride. Usually, anything executed in the blockchain environment is refereed as being \textit{on-chain}, while anything that runs outside the blockchain is referred as \textit{off-chain}. All the operations required to setup ZKSM protocol, as well as the cloaking of trip data, encryption of the offers and generalization of the trip request data, are made off-chain. Whereas the execution of the B-Ride smart-contract and the related transactions are performed on-chain.  

In the following, we evaluate the cost of the different off-chain and on-chain operations.
\subsubsection{Ethereum Baselines and Performance Metrics}

Ethereum introduced the concept of \textit{gas} to quantify the cost associated with each transaction. The cost is payable using the native Ethereum currency, named \textit{Ether}.  Each operation in a smart contract has a fixed cost. For instance, the addition of two variables requires 3 gas, multiplication costs 5 gas has and computing a SHA3 hash needs 30 gas plus 6 gas for every 256 bits of input~\cite{wood2014ethereum}. Therefore, to evaluated the cost of the on-chain operations, we are interested in the following metrics.

\begin{itemize}
\item \textit{Transaction cost:} Is based on the overall gas cost of sending data to the blockchain. Typically, there are four items which make up the full transaction cost; $(i)$ the base cost of a transaction, $(ii)$ the cost of a contract deployment, $(iii)$ the cost of every zero byte of data or code in a transaction, and $(iv)$ and the cost of every non-zero byte of data or code in a transaction~\cite{wood2014ethereum}.


\item \textit{Execution cost:} Indicates the portion of gas that is actually spent on running the code included in a transaction by the Ethereum Virtual Machine (EVM) ~\cite{duan2019aggregating}. 


\item \textit{Storage cost:} Denotes the cost of storing the data in blockchain.

\end{itemize}

Previous metrics can be translated to compute direct
monetary cost on both drivers and riders, and hence to assess the practicality of running our platform on public blockchain. 

\subsubsection{Implementation and Results}

We have implemented a smart contract for Algorithm~\ref{alg1} in  Solidity~\footnote{https://github.com/ethereum/solidity}. B-Ride~\footnote{We note that B-Ride contract in Algorithm~\ref{alg1} is coded as a single "factory contract" in which upon receiving a message with the variable arguments from both entities i.e., riders and drivers, will create a new instance of the child contracts Algorithm~\ref{alg2} and Algorithm~\ref{alg:paymentcontract}} is deployed into the public Kovan Etehreum test network~\cite{Kovan}. In B-Ride, verifying the ZKSM needs some intensive calculation in order to run mathematical and cryptographic functions required by the time-locked deposit protocol. However, implementing these functions in a high-level language, such as Solidity, would be costly in terms of gas. To mitigate this issue, we have deployed a \textit{precompiled contract} that implement these functions. By using a precompiled contracts, less gas is required, as the code is not run on the EVM, but rather on the machine hosting the Ethereum client~\cite{galal2018succinctly}. Typically, a precompiled contract has a fixed address and gas price, and can be invoked using the call operation. Table~6 gives the precompiled contracts used in this paper.

\begin{figure}
 \resizebox{\columnwidth}{!}{
\begin{tikzpicture}
\node (tb2) {
		\begin{tabular}{m{1.4cm}m{1.4cm}m{3.1cm}m{3.7cm}}
		\hline
 Operation & Address & Gas cost & Description  \T \B \\ \hline  
 
\textsf{ECADD} & 0x06 &  500 &Elliptic curve addition. \T \B \\ \hline  

\textsf{PAIRING} & 0x08 & 100000 + 80000 $\times k$ & Optimal ate pairing
check. \T \B \\ \hline   

\textsf{ECMUL} & 0x07 &  40000 & Optimal ate pairing
check. \T \B \\
	 \Xhline{3\arrayrulewidth}
	 \end{tabular}};

\node[above=1.5ex,align=center] at (tb2.north) {\large TABLE 6: The precompiled contracts used in B-Ride.};

\node[below=0.2ex,align=left] at (tb2.south) {Note: in the \textsf{PAIRING}, $k$ denotes the number of points or, equivalently, the \\length of the input divided by 192. The other three operations act on the \\ elliptic curve alt\_bn128.};

\end{tikzpicture}
}
\end{figure}

\begin{figure*}[!t]
	\centering
	\subfloat[On the rider side.\label{fig:a}]
	{\includegraphics[width=0.4\linewidth]{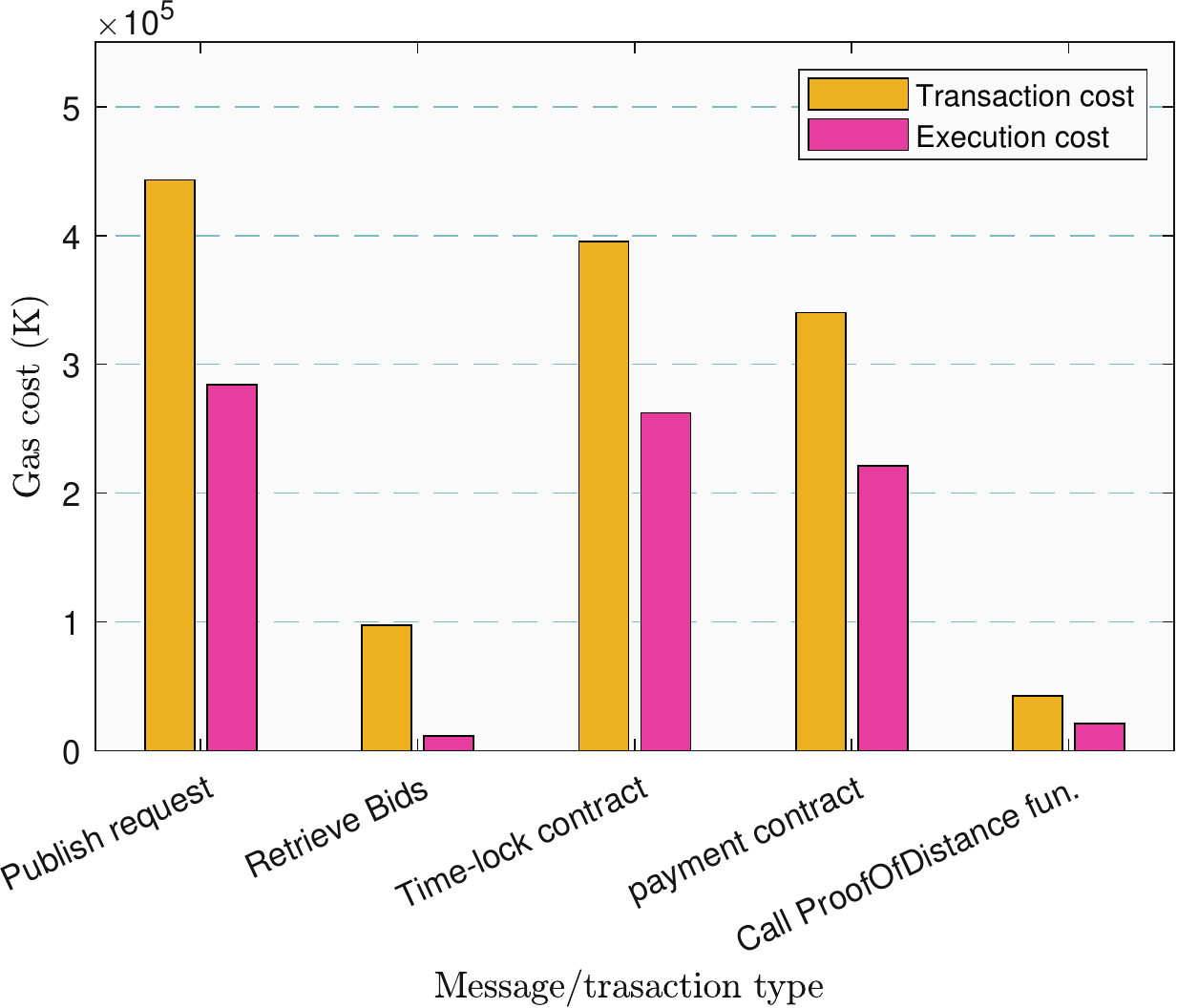}}\hspace{15pt}	
	\subfloat[On the driver side.\label{fig:b}]
	{\includegraphics[width=0.4\linewidth]{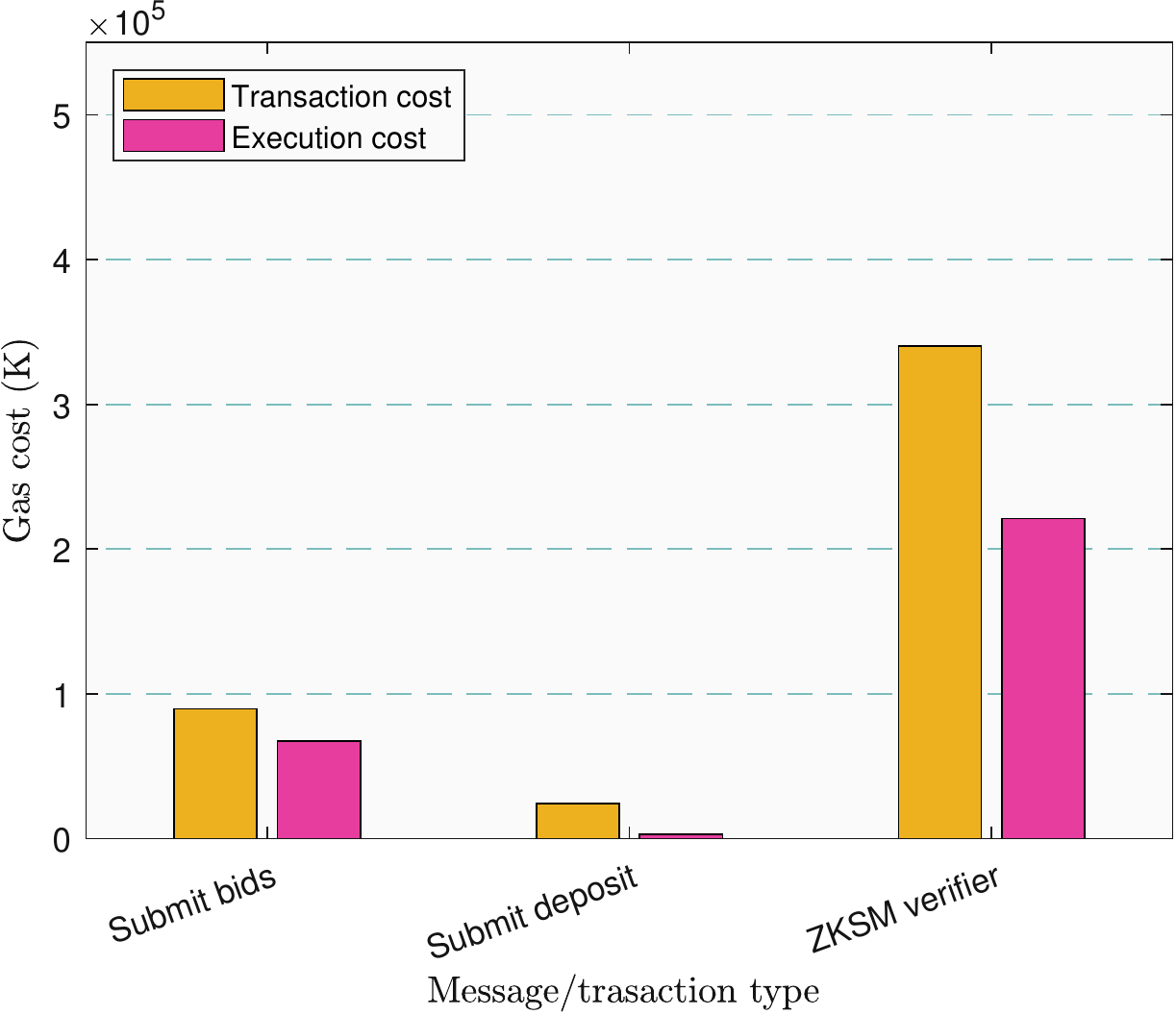}}	
	\caption{The estimated gas cost of calling contract functions on Kovan test net.}
	\vspace{-3mm}
\end{figure*}
\begin{figure}[!t]
	\centering
	\includegraphics[width=.9\linewidth]{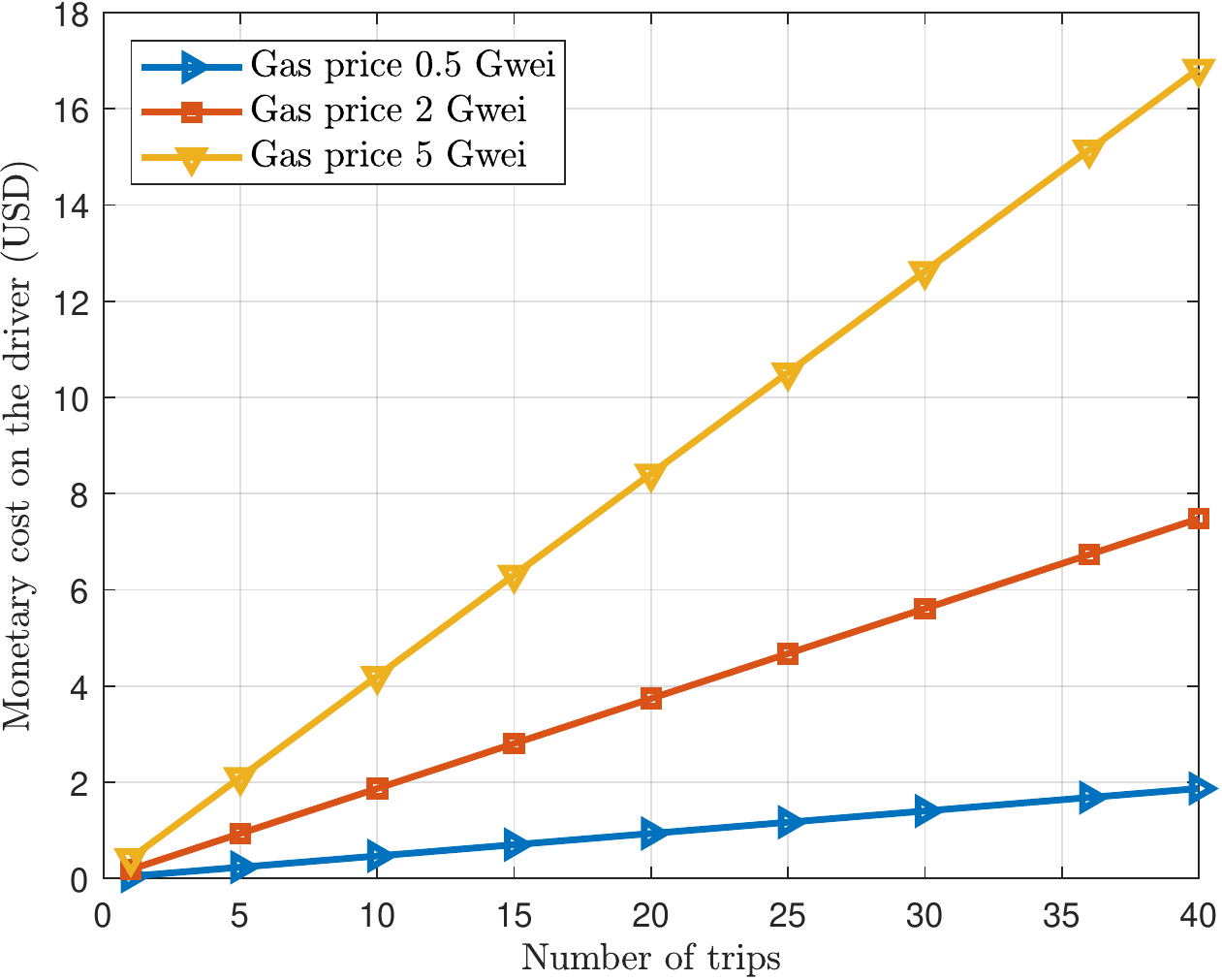}
	\vspace{0mm}
	\caption{The momentary cost of the drivers versus the number of trips.}
	\label{drivercost}
\end{figure}
\begin{figure}[!t]
	\centering
	\includegraphics[width=.9\linewidth]{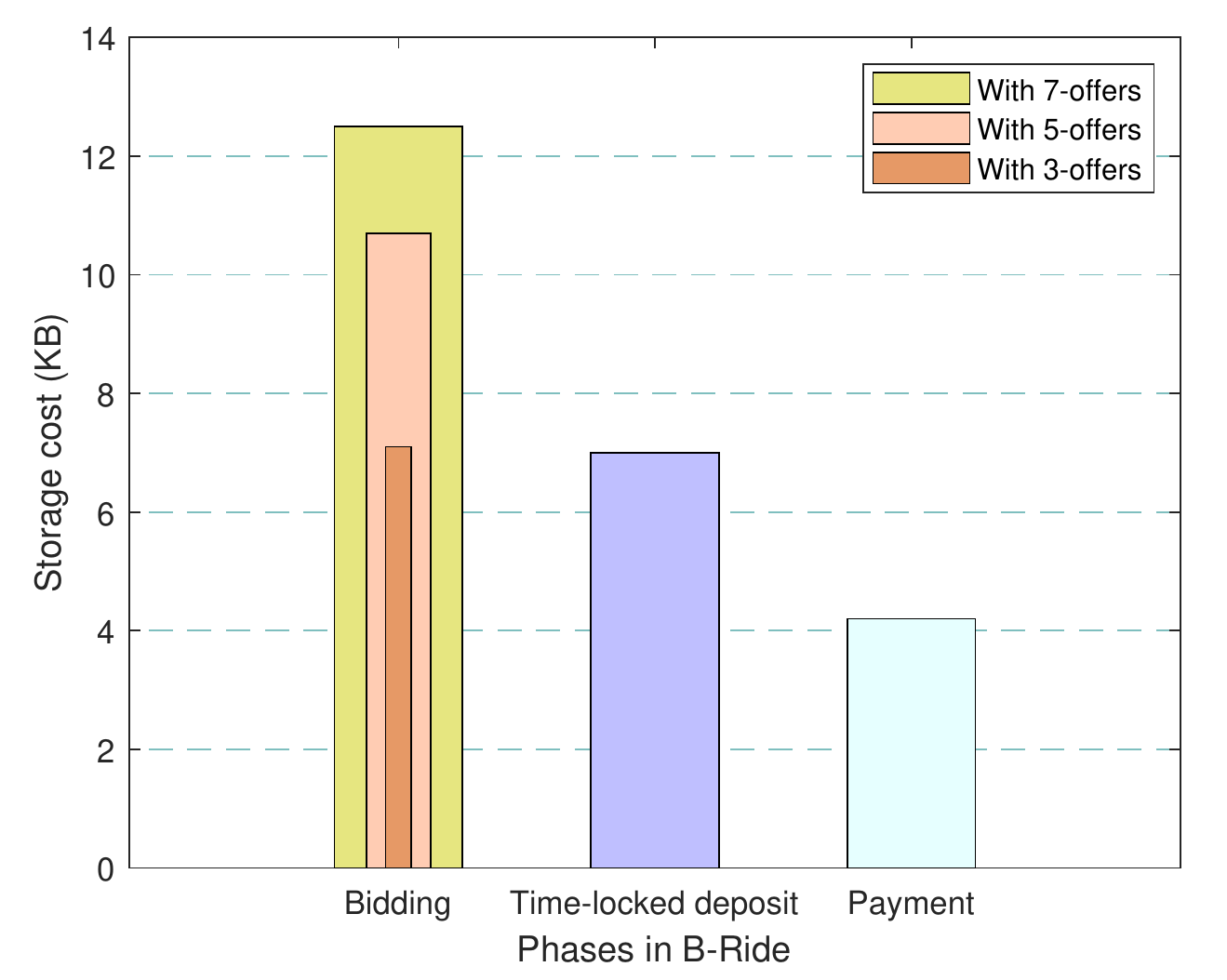}
	\vspace{0mm}
	\caption{On-chain storage cost in B-Ride.}
	\label{storage}
\end{figure}

Fig.~\ref{fig:a} gives the gas consumption per trip costs on the rider side. About 400K gas is required for publishing the bidding contract. Then, 80K gas is required to retrieve the drivers' offers. 320 K gas is needed to deploy the time-locked deposit contract. 340K gas is required for deploying the payment contract, and 42K gas is needed for calling the \texttt{ProofOfDistance} function to perform the fare payment. 

Fig.~\ref{fig:b} reports the gas consumption required by a driver to complete a trip. 89K gas is required to send the encrypted offers to the bidding contract. Once the driver is selected by a rider, 25 Kgas wil be used to send the driver's deposit to the time-locked contract. Finally, the validation of the proof of the underlying ZKSM requires 360K gas.   

Fig.~\ref{drivercost} gives the estimated total cost of driver versus the number of riders, given different gas prices 0.5, 5 and Gwei and Ether price \$217 as of June 25th, 2019\cite{gasstation}. Having 40 trips, the driver costs about \$ 2. By increasing the gas price, the driver costs increase to reach \$ 7.6 for completing 40 trips. The results clearly show that the cost is low and very affordable for the end users.

Fig.~\ref{storage} gives the storage overhead in \textit{Bytes} on the blockchain. Still, the associated storage cost is low and practically acceptable. For instance, with 7 submitted offers, the storage on the blockchain is about only 12 KBytes.

\subsubsection{Off-chain cost}
We evaluate the execution time of the off-chain operations on both the driver and rider side. Raspberry Pi 3 device with 1.2 GHz Processor and 1 GB RAM is used to emulate the driver/rider hardware. We employed BN128 pairing-friendly elliptic curves, that is available on Github~\cite{R15}, to estimate the computation overhead of the ZKSM protocol. The obtained results are summarized in Table 7, and show that the execution time of the off-chain operations takes less than a second. The results demonstrate that the use of ZKSM is adequate for the ride sharing use-case as the additional execution time is within the acceptable application overhead.   

\begin{figure}
\resizebox{\columnwidth}{!}{
\begin{tikzpicture}
    \node (tb2) {
	\begin{tabular}{|m{2.4cm}|m{2.4cm}|m{2.4cm}|}
		\hline
	Operations	& Involved entity & Time (s) \\ 
		\hline	
			ZKSM setup& Rider & 0.5 \\\hline
		ZKSM Proof & Driver & 0.9 \\\hline
	\end{tabular}
	};
    \node[above=1.5ex,align=center] at (tb2.north) {\large TABLE 7: Off-chain overhead.};
\end{tikzpicture}}
\end{figure}

\section{Security and privacy analysis}
\label{sec:analysis}


\smallskip

In this section, we discuss security and privacy concerns \textit{stemmed from decentralizing ride sharing services atop public blockchain} and how B-Ride deal with each of them.

\begin{enumerate}[label={}]
 
 \item \textit{Ensuring correctness and efficiency}. In B-Ride, a rider have a trip while a driver receives the payment of the ride-fare, if and only if they  follow the protocol under the following conditions: $(i)$ The blockchain network can be modelled as an ideal public ledger. $(ii)$ the public key encryption (used in the selection and bidding phase) is correct.  $(iii)$ The underlying ZKSM satisfies completeness.
Regarding \textit{efficiency}, we note the \textit{on-chain} computation in terms of gas consumption and and storage are actually light. This is because of using  precompiled contracts that can validate the proof of the ZKSM by checking a few pairing equalities. 

\smallskip
\item  \textit{Ensuring data confidentiality of riders' activities.} In B-Ride, all related public records of a rider are $(i)$ the  ciphertexts of drivers offers $\{\mathcal{C}_1,\ldots,\mathcal{C}_n\}$, $(ii)$ the ZKSM proof $\pi$, and $(iii)$ the elapsed distances of the trip. The ciphertexts are simulatable due to the semantic security of the public key encryption, and also the proof $\pi$ can be simulated without knowing the secret value (i.e., the pick-up location of the rider) due to the \textit{zero-knowledge} property of the underlying zero knowledge set membership. For the elapsed distance, it cannot reveal any information since it is explicitly known to the rider and the driver.

\smallskip


\smallskip

\item  \textit{Preserving riders' anonymity.} An adversary has three ways to break rider's anonymity: $(i)$ link requests/offers of a driver/rider through his/her blockchain addresses; $(ii)$ link rider/driver through the proofs of arrival sent by the driver; and $(iii)$ link driver/rider in the payment phase.
The first case is trivial, just because the rider will interact with a randomly generated \textit{one-request-only} blockchain address as well as the corresponding public key. For the second case, the anonymity of the rider can be stemmed from the zero-knowledge property of the underlying ZKSM. Besides, the rider can change the set elements of the ZKSM each time he/she looks for a trip. This is allowed due to the feature of the ride-sharing model in which the rider can walk to reach the driver. The last threat is mitigated since only the elapsed distances of the trip  public and no other information about the rider destination is leaked to others. In all three cases, the riders' anonymity is preserved.

\smallskip
\item  \textit{Ensuring security against a malicious driver.} The ways that malicious drivers can cheat are: $(i)$ submitting multiple offers to deliberately let riders have fake reserved trips to make the system unreliable; $(ii)$ providing a valid proof to pick-up location to claim the rider' deposit while not starting the trip with him; and $(iii)$ unfair payment. The first threat is prohibited since the driver has to add a deposit to the time-locked deposit contract within a specific time defined by the smart contract. If the driver does not provide a proof of arrival to the pick-up location, he/she would lose his deposit. The second threat is mitigated by the soundness of the underlying zkSM, which means any incorrect instruction to pass the verification in the smart contract, directly violates the proof-of-knowledge. Moreover, our system builds a reputation system for the driver to indicate their commitment and good willing to complete the planned trips. In the worst case scenario where a malicious driver gets the trip deposit and does not complete the trip, the driver's reputation score of arriving at the pick-up location will increase significantly with his/her reputation score of completed trips.  
 The third threat is handled by allowing the rider to check the distance provided by the driver before the actual fare is paid.

\smallskip
\item  \textit{Ensuring security against a malicious rider.} A malicious rider misbehave by three ways: $(i)$ submit multiple requests to intentionally make fake reservations; $(ii)$ cheat in the ZKSM set-up phase by providing a set $\phi$ with fake signatures of the set elements in the ZKSM in order to deter the driver from claiming his/her trip deposit; $(iii)$ cheat in \textit {payment} phase by providing a forged elapsed distance that is not actually travelled. The first case is mitigated since a driver accepts to share a trip with a rider only if the later adds a ride-deposit to the time-locked deposit smart contract. The second threat is prevented because the time-locked smart contract is public, and the driver can validate the ZKSM set-up using the digital signatures. The last issue is mitigated due to the public blockchain that enables the driver to check whether the rider provides a valid elapsed distance to the payment contract.


 \end{enumerate}

\section{Related work}
\label{Related}
Ride-sharing received a lot of attention in the literature~\cite{asghari2016price,Lightride}. Security and privacy in the centralized setting (client-server model) of ride-sharing service is discussed in~\cite{pham2017oride,rigby2013opportunistic}. Some companies starts developing a blockchain based ride sharing platform e.g., DACSEE~\cite{DACSEE} Arcade City~\cite{ArcadeCity} but without location privacy or anonymity. In the following, we review some of the existing solutions.

In the centralized-setting, in \cite{aivodji2018sride}, SRide has been proposed to address the matching between between drivers and riders in ride-sharing systems while protecting the privacy
of users against both the service provider and curious users. The service provider uses a filtering protocol based on homomorphic arithmetic secret sharing and secure two-party equality
test to determine the subset of drivers with whom the rider can travel.  
     
In~~\cite{sanchez2016co}, a decentralized ride sharing scheme has been proposed. The idea is to use a peer to peer ride-sharing management network instead of the service provider where the peers are the drivers and riders themselves. However, The scheme does not preserve the riders' privacy since the driver and the passenger(s) whose trips match should learn each other's real identity. By linking identities, the rider can be tracked over time and too much information about a specific rider can be obtained. Moreover, the scheme lacks transparency that is provided by the blockchain in B-Ride.


In~\cite{yuan2016towards}, a general blockchain-based intelligent transportation framework is proposed. Also, a case study is presented to discuss the impact of using blockchain in real-time ride-sharing services. Semenko et al.~\cite{semenko2019distributed} proposed a distributed platform for ride-sharing services. The authors suggest having an overlay network that includes all ride-sharing agencies called service nodes to constitute the network layer. The service node is responsible for matching drivers with riders. However, the platform needs trusted infrastructure to run it, and hence may fail in case of one service node is compromised which in turn will lead to inherent problems in the client-server model. Meng et al.~\cite{li2018efficient} proposed a ride-sharing scheme using vehicular fog computing. Road Side Units (RSUs) installed at roads enable local matching between riders and drivers. Anonymous authentication is used to authenticate users while recovering malicious users real identities. A private blockchain made of RSUs is presented to record ride-sharing data in an immutable ledger to enable data auditability. However, using limited resources devices such as RSUs to store massive records of ride-sharing data may be impractical especially in urban areas where there is a high demand to ride-sharing services. 
   
\section{Conclusion}
\label{conclusion}

 In this paper, we have proposed to decentralize ride sharing services using the revolutionary public blockchain named B-Ride.  Analysis, and experiments were conducted to evaluate B-Ride. The results indicate that B-Ride is practical in terms of both on-chain and off-chain overheads. Moreover, it shows the practicability to resolve two main objectives in the
use-case of the decentralized ride sharing  atop public
blockchain: one between the transparency and privacy and the other one between the accountability of system users' and anonymity.  The proposed time-locked deposit protocol ensures security against malicious behaviours of both dishonest drivers/riders.  In addition, the proposed reputation management system tracks drivers' behaviour in B-Ride, allowing them to behave honestly in the system. Otherwise, they will not be selected for future trips. Finally, the rider will have a trip and the driver get the fare in a trust-less environment using the pay-as-you-drive methodology.

  \section{Acknowledgement}
  \label{Ack}
  This work was partially supported by the US National Science Foundation under Grant number 1618549. The statements made herein are solely the responsibility of the authors.

\bibliographystyle{IEEEtran}
\bibliography{CC}	
\begin{IEEEbiography}[{\includegraphics[width=1in,height=1.25in,clip,keepaspectratio]{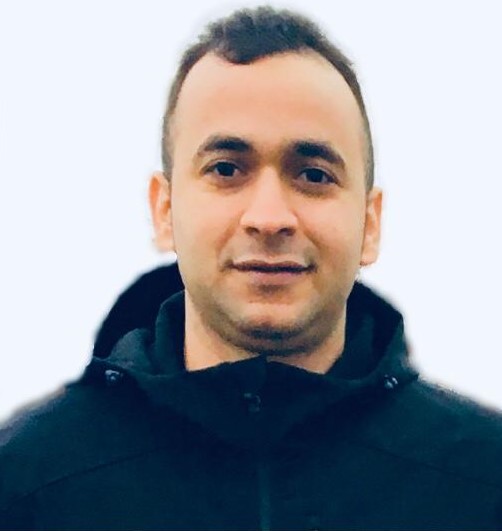}}]{\textbf{Mohamed Baza}} is currently a Graduate Research Assistant in the Department of Electrical \& Computer Engineering, Tennessee Tech. University, USA and pursuing his PhD degree in the same department. He received the B.S. and M.S. degrees in Electrical \& Computer Engineering from Benha University, Egypt in 2012 and 2017, respectively. He has more than two years of experience in information security in Apache-khalda petroleum company, Egypt. His research interests include blockchains, cyber-security, machine learning, smart-grid, and vehicular ad-hoc networks.
\end{IEEEbiography}

\begin{IEEEbiography}[{\includegraphics[width=1in,height=1.25in,clip,keepaspectratio]{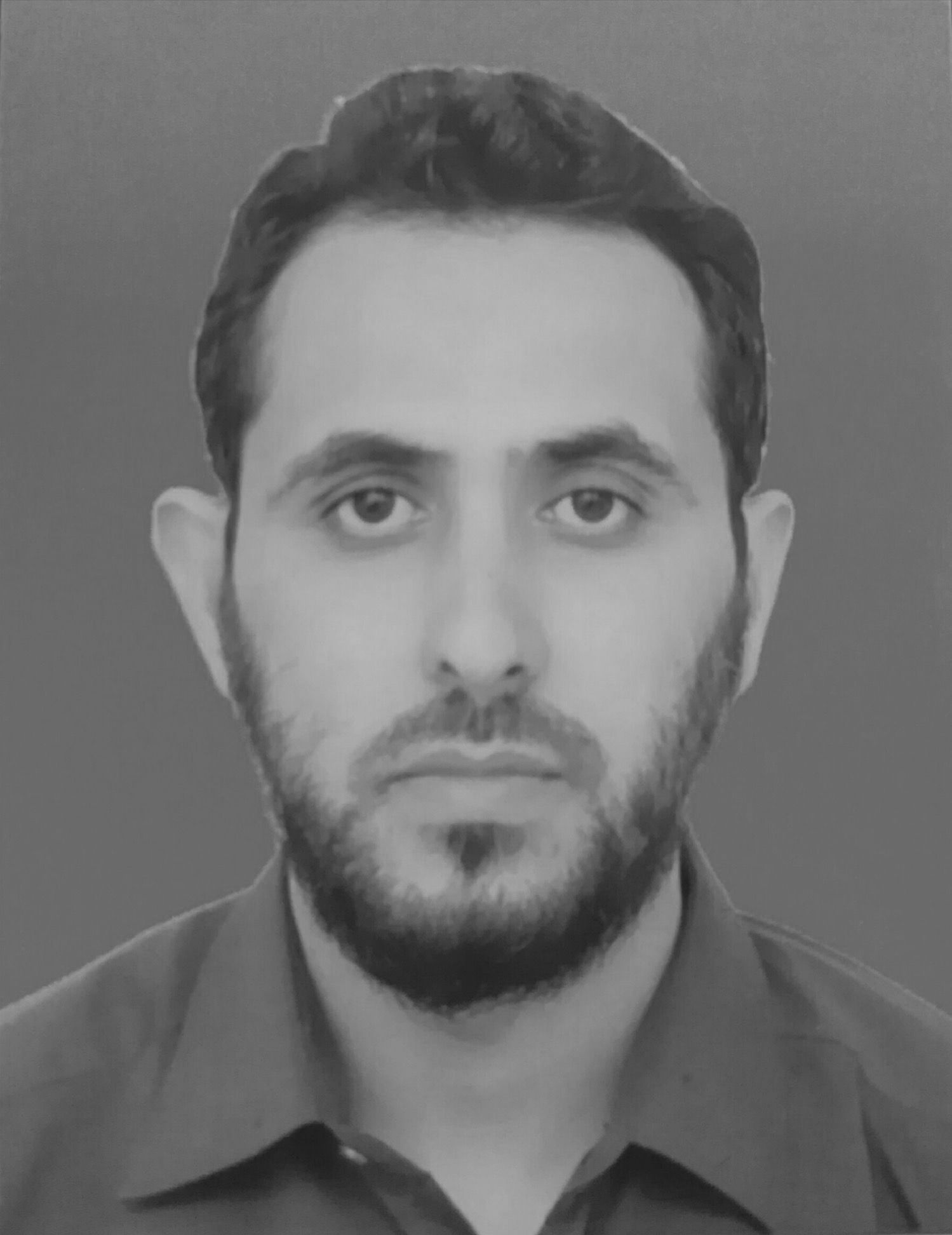}}]{\textbf{Noureddine Lasla}} received the B.Sc. and M.Sc.degrees in 2005 and 2008 from the University of Science and Technology Houari Boumediene (USTHB) and the Superior Computing National School (ESI), respectively, and receivd his PhD degree in 2015 from the USTHB, all in computer science. He is currently a postdoctoral research fellow in the Division of Information and Computing Technology at Hamad Bin Khelifa Univeristy (HBKU), Qatar with expertise in distributed systems, network communication and cyber security.

\end{IEEEbiography}

\begin{IEEEbiography}[{\includegraphics[width=1in,height=1.25in,clip,keepaspectratio]{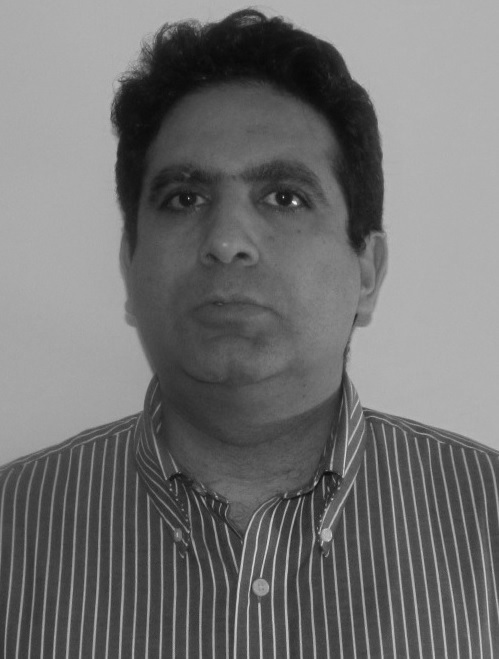}}]{\textbf{Dr. Mohamed M. E. A. Mahmoud}}
received PhD degree from the University of Waterloo in April 2011. From May 2011 to May 2012, he worked as a postdoctoral fellow in the Broadband Communications Research group - University of Waterloo. From August 2012 to July 2013, he worked as a visiting scholar in University of Waterloo, and a postdoctoral fellow in Ryerson University. Currently, Dr Mahmoud is an associate professor in Department Electrical and Computer Engineering, Tennessee Tech University, USA. The research interests of Dr. Mahmoud include security and privacy preserving schemes for smart grid communication network, mobile ad hoc network, sensor network, and delay-tolerant network. Dr. Mahmoud has received NSERC-PDF award. He won the Best Paper Award from IEEE International Conference on Communications (ICC'09), Dresden, Germany, 2009. Dr. Mahmoud is the author for more than twenty three papers published in major IEEE conferences and journals, such as INFOCOM conference and IEEE Transactions on Vehicular Technology, Mobile Computing, and Parallel and Distributed Systems. He serves as an Associate Editor in Springer journal of peer-to-peer networking and applications. He served as a technical program committee member for several IEEE conferences and as a reviewer for several journals and conferences such as IEEE Transactions on Vehicular Technology, IEEE Transactions on Parallel and Distributed Systems, and the journal of Peer-to-Peer Networking.
\end{IEEEbiography}

\begin{IEEEbiography}[{\includegraphics[width=1in,height=1.25in,clip,keepaspectratio]{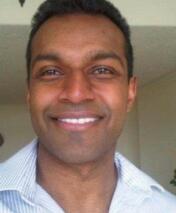}}]{Gautam Srivastava}
was awarded his B.Sc. degree from Briar Cliff University in the U.S.A. in the year 2004, followed by his M.Sc. and Ph.D. degrees from the University of Victoria in Victoria, British Columbia, Canada in the years 2006 and 2012, respectively. He then taught for 3 years at the University of Victoria in the Department of Computer Science, where he was regarded as one of the top undergraduate professors in the Computer Science Course Instruction at the University. From there in the year 2014, he joined a tenure-track position at Brandon University in Brandon, Manitoba, Canada, where he currently is active in various professional and scholarly activities. He was promoted to the rank of Associate Professor in January 2018. Dr. G, as he is popularly known, is active in research in the field of Cryptography, Data Mining, Security and Privacy, and Blockchain Technology. In his 5 years as a research academic, he has published a total of 60 papers in high-impact conferences in many countries and in high-status journals (SCI, SCIE) and has also delivered invited guest lectures on Big Data, Cloud Computing, Internet of Things, and Cryptography at many universities worldwide. He is an Editor of several SCI/SCIE journals. He currently has active research projects with other academics in Taiwan, Singapore, Canada, Czech Republic, Poland, and the U.S.A. He is an IEEE Senior Member and also an Associate editor of the world-renowned IEEE Access journal.
\end{IEEEbiography}

\begin{IEEEbiography}[{\includegraphics[width=1in,height=1.25in,clip,keepaspectratio]{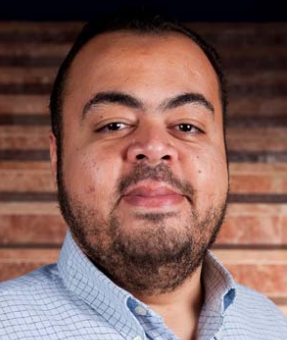}}]{\textbf{Dr.Mohamed Abdallah}} was born in Giza, Egypt. He received the B.Sc. degree with honors from Cairo University, Giza, Egypt, in 1996, and the M.Sc. and Ph.D. degrees in electrical engineering from University of Maryland at College Park, College Park, MD, USA, in 2001 and 2006, respectively. He joined Cairo University in 2006 where he holds the position of Associate Professor in the Electronics and Electrical Communication Department. He is currently an Associate Research Scientist at Texas A\&M University at Qatar, Doha, Qatar. His current research interests include the design and performance of physical layer algorithms for cognitive networks, cellular heterogeneous networks, sensor networks, smart grids, visible light and free-space optical communication systems and reconfigurable smart antenna systems.
\end{IEEEbiography}
\end{document}